\documentclass[unnumsec,webpdf,contemporary,large,namedate]{arxiv}
\usepackage{latexsym}
\usepackage{amssymb}
\usepackage{amsmath}
\usepackage{amsthm}
\usepackage{booktabs}
\usepackage{graphicx}
\usepackage{color}
\usepackage{microtype}
\usepackage{xurl}
\usepackage{hyperref}

\usepackage[utf8]{inputenc}
\usepackage[T1]{fontenc}

\usepackage{amsmath}
\usepackage{amsthm}
\usepackage{amssymb}
\usepackage{amsfonts}

\usepackage{booktabs}
\usepackage{tabularx}

\usepackage{placeins}

\usepackage{wrapfig} 

\usepackage{algorithm}
\usepackage[mathlines, switch]{lineno}

\usepackage{natbib}
\usepackage{cleveref}
\crefname{figure}{figure}{figures}

\makeatletter
\newlength{\mydescwidth}
\newenvironment{description}
  {\list{}{%
      \setlength{\labelwidth}{\mydescwidth}%
      \setlength{\labelsep}{0.75em}%
      \setlength{\leftmargin}{\labelwidth+\labelsep}%
      \setlength{\itemindent}{0pt}%
      \setlength{\topsep}{0.2\baselineskip}%
      \setlength{\partopsep}{0pt}%
      }}
  {\endlist}

\makeatother

\hypersetup{
		pdfencoding=auto, 
		psdextra,
		colorlinks=true,
		citecolor=green!40!black,
		linkcolor=red!50!black,
		urlcolor=blue!80!black
	}

\let\oldemph\emph
\renewcommand{\emph}[1]{\textcolor{blue!60!black}{\oldemph{#1}}}

\usepackage{mathtools}
\usepackage{pifont}
\usepackage{multirow}
\usepackage[export]{adjustbox}
\usepackage{xargs}
\usepackage{xcolor}

\usepackage{booktabs}
\usepackage{nicefrac}
\usepackage{chngpage}
\usepackage{todonotes}
\usepackage{cleveref}
\usepackage{placeins}

\usepackage{tikz}
\usetikzlibrary{cd}

\theoremstyle{remark}

\theoremstyle{plain}

\numberwithin{equation}{section}

\usepackage{todonotes}

\makeatletter 
\def\WFfill{\par 
    \ifx\parshape\WF@fudgeparshape 
    \nobreak 
    \ifnum\c@WF@wrappedlines>\@ne 
    \advance\c@WF@wrappedlines\m@ne 
    \vskip\c@WF@wrappedlines\baselineskip 
    \global\c@WF@wrappedlines\z@ 
    \fi 
    \allowbreak 
    \WF@finale 
    \fi 
} 
\makeatother

\newcolumntype{R}{>{\raggedleft\arraybackslash}X}
\newcolumntype{Z}{>{\centering\arraybackslash}X}
\newcolumntype{W}{>{\centering\arraybackslash}m{.75in}}

\begin{document}

\firstpage{1}

%%%%%%%%%%%%%%%%%%%%%%%%%%%%%%%%%%%%%%%%%%%%%%%%%%%%%%%%%%%%%%%%%%%%%%%%%

% Specify your title and author names here:

\title[Electoral Seat Bias and Polarization]{Effect of Electoral Seat Bias on Political Polarization: A~Computational Perspective}

\author{Daria Boratyn}
\author{Dariusz Stolicki}

\authormark{Boratyn and Stolicki}

\address{\orgdiv{Center for Quantitative Political Science}, \orgname{Jagiellonian University}, \orgaddress{\street{Reymonta 4}, \postcode{31-114}, \state{Krakow}, \country{Poland}}}

\corresp{\href{email:daria.boratyn@uj.edu.pl}{daria.boratyn@uj.edu.pl} \\} 
\corresp{\href{email:dariusz.stolicki@uj.edu.pl}{dariusz.stolicki@uj.edu.pl}} 

% Include a short abstract here (100-300 words):
\abstract{
    Research on the causes of political polarization points towards multiple drivers of the problem, from social and psychological to economic and technological. However, political institutions stand out, because -- while capable of exacerbating or alleviating polarization -- they can be re-engineered more readily than others. Accordingly, we analyze one class of such institutions -- electoral systems -- investigating whether the large-party seat bias found in many common systems (particularly plurality and Jefferson–D’Hondt) exacerbates polarization. Cross-national empirical data being relatively sparse and heavily confounded, we use computational methods: an \emph{agent-based Monte Carlo simulation}. We model voter behavior over multiple electoral cycles, building upon the classic spatial model, but incorporating other known voter behavior patterns, such as the bandwagon effect, strategic voting, preference updating, retrospective voting, and the thermostatic effect. We confirm our hypothesis that electoral systems with a stronger large-party bias exhibit significantly higher polarization, as measured by the Mehlhaff index.
}

\maketitle

%%%%%%%%%%%%%%%%%%%%%%%%%%%%%%%%%%%%%%%%%%%%%%%%%%%%%%%%%%%%%%%%%%%%%%%%%

\section{Introduction}

In many contemporary democracies, elections increasingly resemble confrontations between mutually distrustful political camps rather than routine preference aggregation mechanisms. Scholars document rising ideological distance between parties, growing affective hostility between partisans, and the emergence of increasingly homogeneous, internally reinforcing political communities in both established and newer democracies \citep{GidronEtAl20,Reiljan20,BoxellEtAl22,vanderVeen23}. Polarization is thus rapidly becoming a general feature of modern representative politics.

Extensive evidence points towards a significant adverse impact of the increase in polarization on democratic systems. High levels of partisan animosity and social distance between camps are associated with weaker commitment to democratic norms, greater tolerance of rule-breaking by one’s own side, and a heightened risk that political disagreements are reframed as existential conflicts \citep{KingzetteEtAl21,GrahamSvolik20,Orhan22,LittvayEtAl24,CoxEtAl25}. When citizens view their opponents not merely as wrong but as dangerous, they become more willing to support institutional hardball, democratic backsliding, or even political violence, especially when they believe that ``the other side'' would do the same if given the chance \citep{McCoySomer19,WardTavits19,PasekEtAl22,JanssenTurkenburg25}.

Polarized systems also tend to experience other pathologies of democratic governance. Polarization is linked to legislative gridlock and policy instability, as parties find it harder to assemble cross-cutting coalitions or accept compromise settlements \citep{ThurberYoshinaka15,Binder16}. It can erode trust in institutions, widen perceived gaps in democratic legitimacy between winners and losers, and weaken the informal norms of mutual toleration and forbearance that sustain constitutional democracy \citep{Uslaner15,LevitskyZiblatt18,CoxEtAl25}. At the societal level, affective polarization disrupts social ties, fosters selective exposure to like-minded information, and narrows the set of interlocutors with whom citizens are willing to engage \citep{IyengarEtAl19,FrimerEtAl17,BerntzenEtAl24}. The normative stakes are therefore substantial: in the long run, persistent polarization may undermine both the effectiveness and the perceived fairness of democratic governance.

Research on the causes of political polarization points to multiple drivers of the problem, from social and psychological \citep{DixitWeibull07,Harteveld21,KaplanEtAl22} to economic and technological \citep{McCartyEtAl06,Tornberg22,KubinvonSikorski21}. Among them, political institutions deserve particular attention because, while capable of exacerbating or dampening polarization \citep{BernaertsEtAl23,McCartyEtAl09,Grose20}, they are also, unlike deep social cleavages or technological change, relatively amenable to reform. Electoral systems are especially important because they not only structure how votes are translated into political power, but also affect the number of parties, political fragmentation, government stability and responsiveness, and political incentives \citep{Rae67,Lijphart94a,ShugartTaagepera17}. It is therefore plausible that electoral engineering can mitigate or exacerbate polarization by changing the psychological incentives facing parties and voters \citep{BrunellGrofman05,BesleyPreston07,CarsonEtAl07,McCartyEtAl09}.

In this article we focus on one particular property of electoral systems: \emph{electoral seat bias}. We define it as a systematic advantage for larger parties and corresponding disadvantage for smaller ones in the conversion of votes into seats \citep{SchusterEtAl03,Pukelsheim17,BoratynEtAl25b}. We argue bias is a promising but underexplored institutional mechanism linking electoral systems to polarization, since it affects the strategic landscape faced by both voters and parties: it changes which parties are viable vehicles for representation, modifies the ``cost'' of supporting small or ideologically niche parties, and shapes expectations about how much policy influence an additional vote for a given party is likely to have. In particular, large-party bias gives rational voters incentives to support large over small parties, and numerous studies document that voters respond to such incentives through the strategic desertion of unviable options \citep{CoxShugart96,Spenkuch13,HerrmannEtAl16}. The systematic under-representation of small parties -- which often include moderates, new entrants, or cross-cutting issue entrepreneurs -- may, in turn, compress partisan competition into a narrower set of large parties, fostering sharper inter-group boundaries.

The second mechanism is a \emph{concentration effect}. When electoral rules systematically reward larger parties with bonus seats, votes for larger parties are converted into legislative seats more efficiently than votes for smaller ones. In terms of indirect policy influence, they are ``worth more'' on average. Anticipating this, voters who care primarily about policy outcomes rather than expressive representation acquire an additional incentive to coordinate on one of the leading parties \citep{CoxShugart96}. Over repeated elections, this dynamic can reinforce the emergence of two or three dominant parties or blocs. Party strategists, in turn, may find it electorally profitable to sharpen partisan contrasts -- e.g., by emphasizing identity-laden issues or adopting more distinctive ideological profiles -- since cost of party switching increases as the number of alternatives decreases.

Taken together, threshold and concentration effects suggest a clear hypothesis: \emph{caeteris paribus, electoral systems with stronger large-party seat bias should produce higher equilibrium levels of political polarization among voters.} Yet testing this claim empirically is difficult. Cross-national comparisons are hampered by the endogeneity of electoral rules to existing party systems, by unobserved differences in social cleavages and media environments, and by the fact that seat bias often co-varies with other institutional features. To circumvent these inferential hurdles, we adopt a computational approach. Building on spatial models of party competition and empirically grounded findings on voter behavior, we develop an agent-based Monte Carlo simulation of repeated elections in which we can systematically vary the degree of large-party seat bias while holding other parameters constant. This framework allows us to assess whether, and under what conditions, seat-biased electoral systems indeed tend to generate more polarized voter distributions than more proportional alternatives.

\subsection{Literature review}
\subsubsection{Causes of polarization}

The literature distinguishes analytically separable though intertwined dimensions of polarization. At the elite level, \citet{McCartyEtAl06} and \citet{vanderVeen23} document a long‐run increase in ideological distance between major parties. At the mass level, evidence points to greater ideological consistency among the politically engaged and rising \emph{affective} polarization (more negative feelings toward opposing partisans) \citep{IyengarEtAl12, IyengarEtAl19}. Other work emphasizes that much of the change reflects party sorting -- aligning of partisan identities with ideology -- rather than wholesale movement to the extremes \citep{FiorinaAbrams08,Levendusky09,Mason15,Harteveld21}.

Research converges on the view that polarization is a multi-causal phenomenon \citep{McCoySomer19,BaldassarriPage21}. At the core of the problem are deeply-rooted psychological propensities towards group identity and in-group favoritism \citep{IyengarEtAl12,Mason18,JostEtAl22}, confirmation bias \citep{TaberLodge06}, and cognitive parsimony \citep{KawakatsuEtAl21}. Partisan attachments become social identities bundling ideological, religious, and cultural commitments \citep{Mason18}. Cognitive parsimony drives voters, faced with multiplicity of issues, to follow partisan cues, reinforcing the bundling effect \citep{KawakatsuEtAl21}. Under such conditions, motivated reasoning, in‐group bias, and out‐group animus reinforce one another, contributing to both ideological and affective polarization \citep{IyengarEtAl19,JostEtAl22}. Party sorting and elite cue‐taking further strengthen these dynamics, as citizens align their expressive partisan identities with the positions of national party elites \citep{Levendusky09}.

This is exacerbated by political actors, who deliberately adopt polarization-amplifying strategies because they reap political benefits \citep{GentzkowEtAl19,IyengarEtAl19,RathjeEtAl21}. Some scholars also point to structural social and economic cleavages, especially high and persistent inequality, which provide fertile ground for the formation of antagonistic political blocs \citep{McCartyEtAl06,EstebanRay11,GethinEtAl22}. Moreover, changes in the information environment, including mass adoption of social media, amplify polarizing dynamics by creating an ``echo chamber'' effect and enabling targeted mobilization \citep{HalberstamKnight16,LelkesEtAl17,Levy21,GuessEtAl23,AllcottEtAl24}.

Finally, internal political pressures push political actors towards ideological polarization, despite convergence to the median voter being electorally optimal \citep{Black48,Downs57}. Party platforms, strategies, and nominations are shaped by professional politicians and activists who are more ideologically extreme than the average voter \citep{May73,FiorinaAbrams08,Masket11a,CollittHighton21}. Many of them prefer and emphasize mobilization-based rather than persuasion-based electoral strategies \citep{FisherDenver09,Panagopoulos16}, which pull parties away from the center and toward the activist base \citep{Aldrich83,LaymanEtAl06,Amitai24}. Moreover, strongly ideological individuals speak out more than moderates, increasing polarization in public debate \citep{BurnettEtAl22}. Social media algorithms reinforce this by prioritizing emotionally engaging content \citep{BradyEtAl17}. This environment makes extreme views seem more widespread than they are, further encouraging shifts away from the center \citep{WestfallEtAl15}.

Most sources of polarization are hard to remedy. However, political institutions are an exception here. Comparative research shows that institutional context, and particularly the electoral system, shapes how social cleavages are translated into party competition and whether conflict is aggregated into a few large blocs or dispersed across multiple actors \citep{Lijphart94a,Lijphart99}. At the same time, institutional re-engineering, while not without its own challenges, is among the most feasible remedies to polarization. The next subsections therefore turn to the literature on the political consequences of electoral systems and on electoral seat bias in particular.

\subsubsection{Political effects of electoral systems}

Classic work argues that electoral law features such as district magnitude, formula, and legal thresholds shape the format of the party system, political incentives, and the prospects for single-party versus coalition governments \citep{Rae67,Lijphart94a,TaageperaShugart89,Gallagher92}. Mechanical effects (how votes translate into seats) and psychological effects (how voters and parties adapt their strategies to those rules) jointly generate familiar regularities such as the Duverger and micromega laws \citep{Duverger51,Colomer04}.

A smaller but growing body of work examines how electoral institutions intersect with polarization. Some comparative studies suggest that proportional systems may facilitate the representation of diverse interests and thereby dampen zero‐sum, two‐bloc conflict, whereas majoritarian systems tend to align competition along a single dominant divide \citep{Lijphart99}. Others argue effects are conditional: permissive rules in unequal societies may yield polarized party offerings, while restrictive rules compress conflict into fewer, sharper blocs \citep{Han15}. Yet, when considering polarization, research often focuses on broad system types (e.g., majoritarian versus proportional) rather than specific rule properties such as seat bias.

\subsubsection{Seat bias in electoral systems}

Seat bias is the expected gap between a party’s vote share and actually allocated seat share. For apportionment (proportional) methods, asymptotic seat biases can be analytically derived under various distributional assumptions \citep{SchusterEtAl03,Janson14,Pukelsheim17,BoratynEtAl25b}. These studies show that divisor methods with lower rounding thresholds (such as Jefferson–D’Hondt) tend to be large‐party biased, whereas those with higher thresholds (e.g., Sainte–Laguë or Adams) are more neutral or even small‐party biased.

Recent work extends seat-bias analysis to multi-district contexts. A relation between seat bias and mean district magnitude has been already noted by \citet{TaageperaShugart89} and \citet{Rae95}. For stationary divisor methods, \citet{FlisEtAl20} and \citet{BoratynEtAl25b} established the precise functional form of that relation, confirming that large-party bias increases with mean district magnitude and the number of competing parties. For systems employing single-member districts, bias depends primarily on the geographic distribution of voters. In particular, the more concentrated the distribution of district-level results, the greater the large-party bias of the plurality (FPTP) rule.

\section{Methods}

Establishing a causal link between electoral seat bias and political polarization using observational data is challenging. Cross-national comparisons face the usual problems of comparative research: institutions co-vary with social cleavages, party-system histories, media systems, and economic development, while electoral rules are often endogenous to earlier political conflict. Even within-country reforms rarely generate clean ``natural experiments'' in which large-party bias changes while other features of the political environment remain stable. Empirical studies that correlate institutions with polarization are therefore vulnerable to omitted-variable bias, reverse causality, and contextual heterogeneity.

Computational methods offer a useful alternative to such empirical approaches. In particular, the ``generative'' approach of agent-based modeling enables us to construct artificial electorates by specifying micro-level behavioral rules and institutional contexts which, when iterated, reproduce macro-level regularities of interest \citep{Epstein99}. We can develop models in which the electoral system parameters are exogenously set and all other features are held fixed, thus isolating the effect of electoral seat bias on polarization by systematically varying the relevant institutional parameters while controlling, by design, for confounding factors \citep[see also][]{deMarchi05,MacyFlache11}.

The agent-based approach is particularly well suited for analysis of voter behavior and party competition. Elections are a complex adaptive system in which individual-level choices, heuristics, and social influences interact with institutional incentives to produce emergent macro-level patterns -- such as party fragmentation, volatility, and polarization -- that may be highly non-linear and path dependent \citep{MacyFlache11}. Analytical models of electoral competition, especially in multi-party spatial settings with strategic voting and dynamic preference updating, quickly become intractable. Agent-based models have therefore been increasingly used to study party competition and electoral dynamics in multi-dimensional policy spaces, demonstrating how simple behavioral rules at the micro level can generate rich and sometimes counterintuitive aggregate behavior \citep{LaverSergenti11}.

In this spirit, we employ an agent-based Monte Carlo simulation to study how electoral seat bias shapes the long-run evolution of mass-level polarization. Our model specifies a population of voters and parties embedded in a spatial policy environment, a set of decision rules governing vote choice and preference updating, and a family of electoral formulas characterized by different degrees of large-party bias. By repeatedly simulating electoral cycles and varying only the institutional parameters of interest, we obtain distributions of polarization outcomes under alternative electoral regimes. The next subsection details the design of this simulation, including the initial state of the party system and voter preferences, the behavioral rules governing vote choice and preference change, and the formal representation of electoral systems with varying seat-bias properties.

\subsection{Simulation design}

We start with a traditional \emph{spatial model}: every voter and every party is assigned a position (ideal point) in a two-dimensional policy space \citep{GoodTideman76,EnelowHinich84}. We assume the policy space to be spherically-symmetrical (unlike most expert survey datasets, such as Chapel Hill Expert Survey, \citealp{BakkerEtAl15}, which implicitly assume a cubical policy space), as this is better supported by empirical research: ideal points obtained through multidimensional scaling of data sources such as roll-call \citep{PooleRosenthal91,Poole05,HixEtAl06}, textual \citep{KimEtAl18,BoratynEtAl23a}, and social media data \citep{HemphillEtAl16} tend to approximate spherical rather than cubical distributions. While for NOMINATE data this may be an artifact of its use of principal component analysis \citep{Potthoff18}, other scaling methods, free of such bias -- such as Bayesian methods \citep{ClintonEtAl04} and UMAP \citep{McInnesEtAl18} -- also identify approximately spherical distributions \citep{Berggren17,McInnes18}. Analyses of congressional voting also point towards distances across the two principal dimensions being negatively correlated \citep{ChatterjeeEyigungor23}. Moreover, a spherical policy space is mathematically convenient: it is rotationally invariant, thus avoiding an arbitrary commitment to a particular basis, and agrees very well with the assumption that voter preferences are Gaussian.

Accordingly, \emph{party ideal points} are drawn from a uniform distribution on a circle of radius $r$ centered at the origin. Initial \emph{voter ideal points} are a mixture of several distributions: with probability $3/4$ a voter is an initially uncommitted voter, with an ideal point drawn from a multivariate Gaussian distribution centered at the origin, while with probability $1/4$ a voter is a \emph{party base voter}. Party base voters are distributed among parties in proportion to initial party sizes, and their ideal points are drawn from a multivariate Gaussian distribution centered at their party's ideal point.

This model reflects three basic assumptions. First, we assume the electorate to be a mixture of voters with strong partisan preferences and swing voters, who can be more easily swayed by any party. Omitting the former would be unrealistic and could lead to parties clustering too strongly in the center, while omitting the latter would lead to a gross underestimation of electoral volatility. Second, we expect voter preferences within each group to be approximately normally distributed \citep{Rice28,Granberg87,FowlerEtAl23,BertsouEtAl25}. Third, however, we assume that parties are more likely to be ideologically expressive than voters \citep{BafumiHerron10} and more likely to cover most of the issue space -- hence the choice of the uniform rather than Gaussian distribution.

The \emph{initial party sizes}, in turn, are obtained from the vector of expected magnitudes of $i$-th largest party calculated under the assumption that the vector of party sizes is uniformly distributed on the unit simplex \citep{BoratynEtAl20}. It is, however, flattened by taking the square root to better match empirical distributions of party sizes. Furthermore, in light of the findings of \citet{Taagepera07} that the theoretically expected sizes of the largest and second-largest parties are, respectively, inflated and deflated in comparison to the empirical data, we artificially level both by averaging them.

At the stage of simulating individual voter behavior, however, we depart from the traditional spatial model. While our voters still prefer parties that are closer to their ideal points to those that are more distant, this is not the only criterion determining their choices. We also incorporate several voting patterns known from empirical research on electoral behavior:
\begin{itemize}
    \item \emph{strategic voting} \citep{MyersonWeber93,BlaisEtAl01}: voters can prefer a more distant alternative to a closer one in order to minimize the risk of an even more distant party winning an election or of their own vote being wasted; in practice, we expect this pattern to translate into a preference for larger parties over smaller ones;
    \item the \emph{bandwagon effect} \citep{LanoueBowler98}, according to which voters rally to the perceived public opinion leaders, again favoring larger parties over smaller ones; and
    \item the \emph{retrospective voting} pattern \citep{Fiorina81}, according to which at least a portion of voters use their vote primarily to reward or punish the governing party, depending on whether they approve or disapprove of its performance in office.
\end{itemize}

Accordingly, our voter decision function is as follows:
\begin{itemize}
\item First, for each voter, we order parties according to the decreasing ratio of the party size, i.e., its number of seats, to the distance between the voter and party ideal points taken to some power $\alpha_j$, which is determined for each voter at random (taken from a truncated normal distribution) and represents individual propensity to prefer policy over strategy: the higher a voter's $\alpha_j$, the less likely that voter is to prefer a larger but more distant party over a smaller but closer one.
\item Second, for dissatisfied voters, the governing party is struck from the preference ranking of parties (in the initial iteration, the set of dissatisfied voters is empty).
\item Finally, each voter chooses the top-ranked remaining party and votes for it.
\end{itemize}

Votes are then counted, and seats are allocated according to the specific electoral system we are simulating. Because we do not simulate a partition of the electorate into electoral districts, for each electoral system we use a ``seats-votes approximation'' -- a mathematical device that enables us to simulate the allocation of seats under a given system using only aggregate data. We discuss seats-votes approximations for the two electoral systems of greatest interest to us -- Jefferson-D'Hondt and single-winner plurality (FPTP) -- later in this section.

At this point, voters adjust their positions in response to the electoral cycle, marking the second major divergence between our simulation and the traditional spatial model. This adjustment is a sum of two \emph{preference updating} processes:

\begin{enumerate}
    \item \emph{Random perturbation of voter preferences.} We assume that voter policy preferences can shift in any direction in response to random life experiences causing value reassessments. However, we expect most such shifts to be small.
    \item \emph{Affective shift.} The basic underlying assumption here is that affective polarization, which in our model is an independent social process, pulls party voters towards that party's position \citep{CallanderCarbajal22}. The magnitude of this effect depends on the voter and varies from $1$ (complete voter identification with the party ideal point), through $0$ (no change in voter position), to negative values (a shift away from the party ideal point). However, we assume that the shift effect distribution is generally skewed towards polarization, and that voters already close to the party position are more likely to get pulled even closer, while voters that are distant are about just as likely to shift towards the party as to shift away from it.
    \item \emph{Thermostatic effect} \citep{Wlezien95}. We observe that effect only for the governing party: voters disapproving of its performance in office move away from its ideal point and towards a competitor that had the second place in their preference ordering. Mechanically, this effect is sort of a mirror of the affective shift effect.
\end{enumerate}

However, before any preferences are updated, voter satisfaction is recalculated. We assume the largest party to be the governing party. The proportion of its voters approving of the government's performance in office is drawn from a beta distribution whose expected value decreases with the number of consecutive terms of office served by the present governing party. Then, each voter's satisfaction with the governing party's performance is drawn from a Bernoulli distribution with the expectation equal to that approval rating.

After those adjustments, we repeat the process, starting with the recalculation of voter preference rankings and votes.

\subsection{Formalization}

\subsubsection{Notation}

Let:
\begin{description}
    \item[$v \in \mathbb{N}_{+}$] be the number of voters,
    \item[$n \in \mathbb{N}_{+}$] be the number of parties, 
    % \item[$s \in \mathbb{N}_{+}$] be the number of seats to be allocated,
    % \item[$c \in \mathbb{N}_{+}$] be the number of districts,
    % \item[$m := s/c$] be the mean district magnitude,
    \item[$m$] be the mean district magnitude,
    \item[$\delta(x)$] be the Dirac measure concentrated at $x$,
    \item[${[l]}, l \in \mathbb{N}_{+}$] denote the set $1, \dots, l$.\smallskip
\end{description}

\noindent Moreover, we use the letter $q$ to denote vote shares, the letter $s$ to denote seat shares, the letter $i$ to index parties, the letter $j$ to index voters, the letter $k$ to index iterations, and uppercase letters to denote random variables.

Finally, let
\begin{equation}
\Delta_{n} := \{\mathbf{x} \in \mathbb{R}_{\ge 0}^{n}: \sum_{i=1}^{n} x_i = 1\}
\end{equation}
denote the $(n-1)$-dimensional unit simplex.

\subsubsection{Simulation parameters}

Let:
\begin{description}
    \item[$r \in \mathbb{R}_{+}$] be the radius of a disc containing party positions,
    \item[$\sigma \in \mathbb{R}_{+}$] be the standard deviation of the distribution of initial voter positions, 
    \item[$\rho \in \mathbb{R}_{+}$] be the scaling factor for the distribution of `party base' voters,
    \item[$\mu \in \mathbb{R}_{+}$] be the location parameter of the truncated normal distribution from which $\alpha_j$ will be drawn,
    \item[$\tau \in \mathbb{R}_{+}$] be the standard deviation parameter of the truncated normal distribution from which $\alpha_j$ will be drawn,
    \item[$\beta_1, \beta_2 \in \mathbb{R}_{+}$] be the parameters of the beta distribution from which the government job approval will be drawn,
    \item[$\pi \in (0, 1)$] be the probability of a random perturbation of voter preferences,
    \item[$\varsigma \in \mathbb{R}_{+}$] be the standard deviation parameter of the distribution of random preference perturbation,
    \item[$\lambda \in \mathbb{R}_{+}$] be the standard deviation parameter of the lognormal distribution from which voter shift parameter will be drawn.
\end{description}

\subsubsection{Random variables}

For every iteration of the electoral cycle $k$, $k \in \mathbb{N}_+$, every party~$i$, $i = 1, ..., n$, and every voter $j$, $j = 1, ..., v$, let:
\begin{itemize}
    \item $P_i$ be the position of the $i$-th party in the $2$-dimensional policy space, $\mathbb{R}^2$,
    \item $Q_i^k$ be the vote share obtained by the $i$-th party in the $k$-th iteration,
    \item $S_i^k$ be the seat share obtained by the $i$-th party in the $k$-th iteration,
    \item $X_j^k$ be the position of the $j$-th voter in the policy space in the $k$-th iteration,
    \item $W^k$ be the index of the party that is the election winner in the $k$-th iteration,
    \item $J^k$ is the overall job approval of the government following the $k$-th iteration,
    \item $A_j^k$ be a Boolean variable that equals $1$ if and only if the $j$-th voter approves of the performance of the governing party in the $k$-th iteration.
\end{itemize}

\subsubsection{Initial conditions}

For every $i$ we assume that $P_i$ (ideal point of the $i$-th party) is a random variable with the uniform probability distribution on a disk of radius $r$, i.e.,
\begin{equation}
P_i \sim \mathrm{Unif}\Big(B\big((0,0), r\big)\Big).
\end{equation}

\begin{figure}[htb]
\includegraphics[width=\columnwidth]{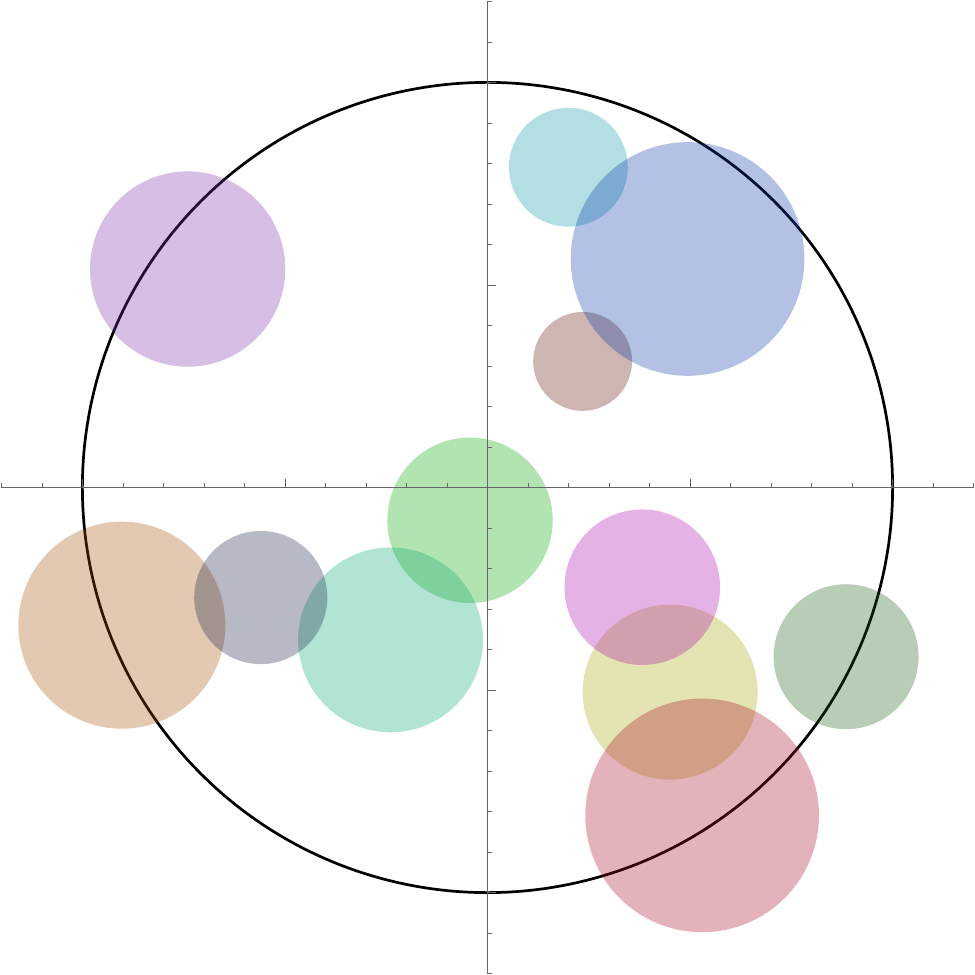}
\caption{Running example: initial party sizes and positions. Parties are represented by colored circles centered at ideal points and with area proportional to initial size. Simulation parameters correspond to those given in the next section.}
\label{fig:step0}
\end{figure}

For every $i$, let $S_i^0$ be the initial size of the $i$-th party. We define it as a function of the expected values of the coordinates of a vector $Z := (Z_1, ..., Z_n)$ drawn from a uniform distribution on an $(n-1)$-dimensional unit simplex $\Delta_n$, i.e., 
\begin{equation}
S_i^0 : = \begin{cases} \frac{\sqrt{\mathbb{E}(Z_1^{\downarrow})}+\sqrt{\mathbb{E}(Z_2^{\downarrow})}}{2\sum_{l=1}^n \sqrt{\mathbb{E}(Z_l^{\downarrow})}}, & i = 1,2, \medskip \\
\frac{\sqrt{\mathbb{E}(Z_i^{\downarrow})}}{\sum_{l=1}^n \sqrt{\mathbb{E}(Z_l^{\downarrow})}}, & i = 3, ..., n,
\end{cases}
\end{equation}
where $Z_l^{\downarrow}$ denotes the $l$-th largest coordinate of $Z$. In other words, we take a coordinate-wise square root of the vector of expected order statistics of $Z$, renormalize, and average the first two coordinates. From \citet{BoratynEtAl20} it follows that $$\mathbb{E}(Z_l^{\downarrow}) = \frac{1}{n} \sum_{k = l}^n \frac{1}{k},$$
wherefore
\begin{equation}
S_i^0 : = \begin{cases} \frac{\sqrt{\sum_{k = 1}^n 1/k }+\sqrt{\sum_{k = 2}^n 1/k }}{2\sum_{l=1}^n \sqrt{\sum_{k = l}^n 1/k }}, & i = 1,2, \medskip \\
\frac{\sqrt{\sum_{k = i}^n 1/k }}{\sum_{l=1}^n \sqrt{\sum_{k = l}^n 1/k}}, & i = 3, ..., n.
\end{cases}
\end{equation}
An example realization of the distribution of party ideal points and initial sizes is plotted on Figure \ref{fig:step0}.

For every voter, that voter's initial position $X_j^1$ is a random variable drawn from a mixture of a multivariate Gaussian centered at the origin (for ``uncommitted'' voters) and multivariate Gaussians (one per party) centered at party ideal points (for ``party base'' voters):
\begin{equation}
X_j^1 \sim \frac{3}{4} \mathcal{N}\big((0, 0), \sigma^2\big) + \frac{1}{4} \sum_{i=1}^{n} S_i^0 \mathcal{N}\big(P_i, (\sigma \rho S_i^0)^2\big).
\end{equation}
The standard deviation for the party base voters is rescaled, first by $\rho$ to ensure that party base voters are more concentrated around the party than the uncommitted voters around the origin, and then by the initial party size to compensate for smaller parties having lesser ``gravitational attraction.''

Finally, since the voter satisfaction effects only come into play into the second and successive iterations, we fix $A_j^0 \sim \delta(1)$ for every voter and $W^0 \sim \delta(0)$.

\newpage
\subsubsection{Iterations of the electoral cycle}

\noindent\textbf{Voting}\smallskip

\noindent For every iteration of the electoral cycle $k \in \mathbb{N}_+$, every party $i = 1, ..., n$, and every voter $j = 1, ..., v$, we define the utility of the $i$-th party for the $j$-th voter as:
\begin{equation}
    U_{i,j}^k := \frac{S_i^{k-1}}{d(X_j^k, P_i)^{\alpha_j}},
\end{equation}
where $d$ is the $L_2$ (Euclidean) metric and $\alpha_j$ is a random variable drawn from a normal distribution left-truncated at 1 with mean~$\mu$ and standard deviation $\tau$:
\begin{equation}
    \alpha_j \sim \mathcal{T}\big(\mu, \tau, (1, \infty)\big).
\end{equation}
Recall that this formula accounts for the strategic component of voting behavior and for the bandwagon effect: voters balance party's ``gravitational attraction,'' which is proportional to its size, with its proximity to their positions. The relative importance of proximity depends on the voter and is drawn from a truncated normal distribution -- we need a left-truncated distribution to ensure that no voter will weigh proximity negatively, but we also desire more flexibility than could be afforded by one-parameter distributions such as exponential.

A $j$-th voter's vote in the $k$-th iteration is usually the index of the party that maximizes their utility in that iteration. However, we also account for the retrospective voting pattern: voters sufficiently disapproving of the governing party's performance will not vote for it even if it would otherwise be an optimal choice for them. Thus,
\begin{equation} \label{eq:vote}
    V_j^k := \operatorname*{arg\,max}_{i \in [n]} \begin{cases}
        U_{i,j}^k & \textrm{ if } i \neq W^{k-1} \textrm{ or } A_j^{k-1} = 1, \\
        0 & \textrm{ if } i = W^{k-1} \textrm{ and } A_j^{k-1} = 0.
    \end{cases}
\end{equation}

Then, we recalculate the vote shares,
\begin{equation}
    Q_i^k := \frac{|\{j \in [v]: V_j^k = i\}|}{v},
\end{equation}
and the seat shares,
\begin{equation}
    (S_1^k, \dots, S_n^k) := f((Q_1^k, \dots, Q_n^k)),
\end{equation}
where $f: \Delta_n \longrightarrow \Delta_n$ is a \emph{seats-votes function} representing the specific electoral formula used for the model.

Finally, we determine the election winner:
\begin{equation}
    W^k := \operatorname*{arg\,max}_{i \in [n]} S_i^k.
\end{equation}

\noindent\textbf{Voter satisfaction}\smallskip

\noindent After the simulated election, voter satisfaction is recalculated. First, the overall job approval of the government, $J^k$, is drawn from a beta distribution with parameters $\beta_1$ and $\kappa \beta_2$, where $\kappa$ is the \emph{cost-of-ruling factor}, i.e., the number of consecutive elections $l = 1, ..., k$ such that $W^k = W^{k-l+1}$:
\begin{equation} \label{eq:jobApproval}
    J^k \sim \operatorname{Beta}(\beta_1, \kappa \beta_2).
\end{equation}
Then, for every voter $j \in [v]$, their approval of the government is drawn from the Bernoulli distribution with parameter $J^k$:
\begin{equation}
    A_j^k \sim \operatorname{Bernoulli}(J^k).
\end{equation}
The cost-of-ruling factor in (\ref{eq:jobApproval}) accounts for the fact that not only the exercise of power alienates some voters, but the effect accumulates with each consecutive term of office the party has been in government \citep{Paldam86,NannestadPaldam00}.

\begin{figure}[tb]
\includegraphics[width=\columnwidth]{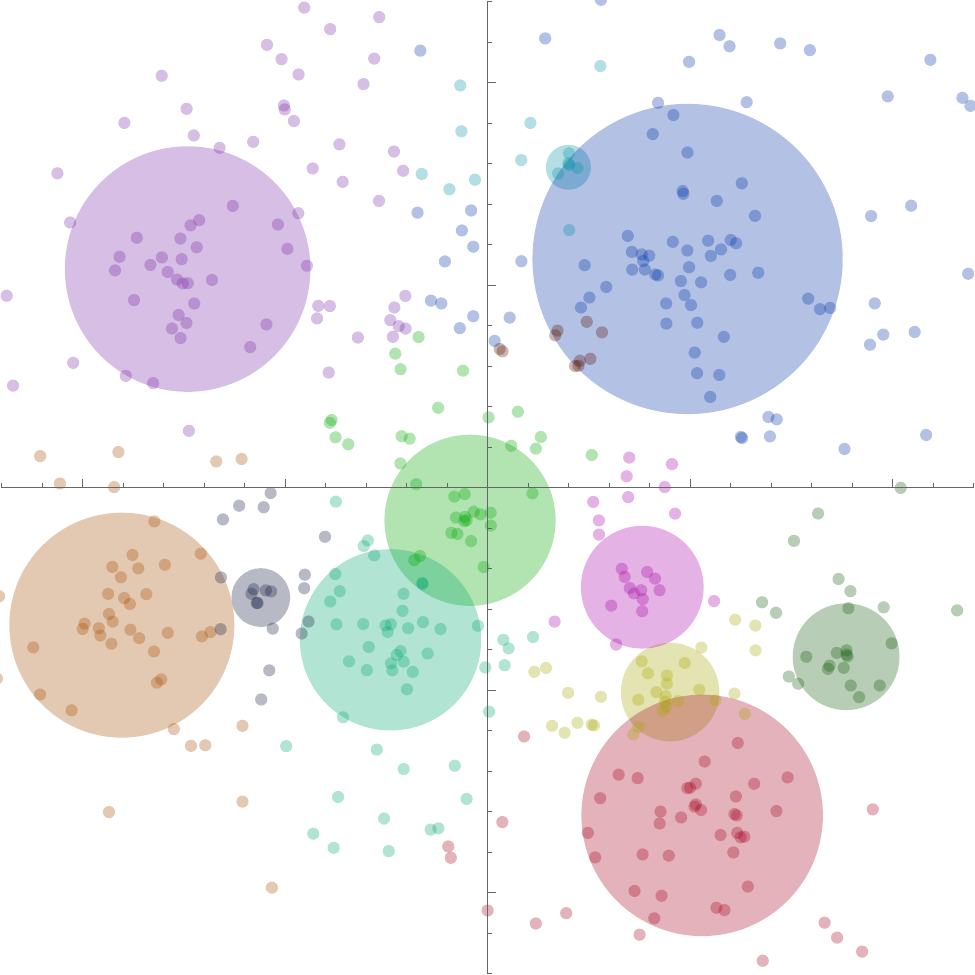}
\caption{Running example: first iteration. Seats were allocated using the Jefferson--D'Hondt method with $m = 12$. Colored circles represent parties after the first election (area of each party is proportional to its seat share), while colored dots represent a sample of voters. Note that there are voters without corresponding party circle -- they have voted for a party that failed to win any seats due to the natural threshold.}
\label{fig:step1}
\end{figure}

\smallskip

\noindent\textbf{Preference updating}\smallskip

Finally, we arrive at the preference updating stage. Each of the updating effects is computed independently. We start with random perturbation, with probability $\pi$ drawn from the multivariate normal distribution centered at the origin and with standard deviation $\varsigma$, and with probability $1 - \pi$ equal to zero:
\begin{equation}
    \delta_j^k \sim \pi \mathcal{N}\big((0,0), \varsigma^2\big) + (1 - \pi) \delta\big((0,0)\big).
\end{equation}
In practice, we expect $\pi$ to be no greater than $1/2$ (to reflect the fact that only a minority of voters is randomly perturbed in each iteration). Similarly, $\varsigma$ should be significantly smaller than $\sigma$, again to reflect the fact that random perturbations are likely to be relatively small.

Simulation of affective shifts is more complex. The shift vector itself is obtained by multiplying a displacement vector between the voter's position and the position of that voter's chosen party by a random magnitude variable in $(-\infty, 1)$. In such a case, a magnitude of $1$ would, in the limit, correspond to the voter's position becoming identical with the party position, a magnitude of $0$ would correspond to no change, and negative magnitudes would mean a shift away from the party position. As noted, since we assume that underlying social trends favor affective polarization, the magnitude distribution needs to be asymmetric and right-skewed. Since there are no canonical probability distributions supported on $(-\infty, 1)$, we define shift magnitude as $1 - \eta_j^k$, where $\eta_j^k$ is lognormally-distributed:
\begin{equation}
    \eta_j^k \sim \mathcal{L}(\lambda \Phi_j^k - 1, \lambda),
\end{equation}
and $\Phi_j^k$ is the image of $d(X_j^k, P_{V_j^k})$ under the probability integral transform $x \mapsto \Pr(d(X_L^k, P_{V_j^k}) \le x)$, wherein $L \sim \operatorname{Unif}\left(\{l \in [n]: V_l^k = V_j^k\}\right)$ is a random voter supporting the same party as voter $j$. Thus, the closer a voter already is to the party position, the more likely he is to experience a strong affective shift towards it. Since being close to or distant from a party can only be understood in relation to how close or distant its other voters are from that party, we normalize the $L_2$ distance through the use of the probability integral transform.

\begin{figure}[htb]
\includegraphics[width=\columnwidth]{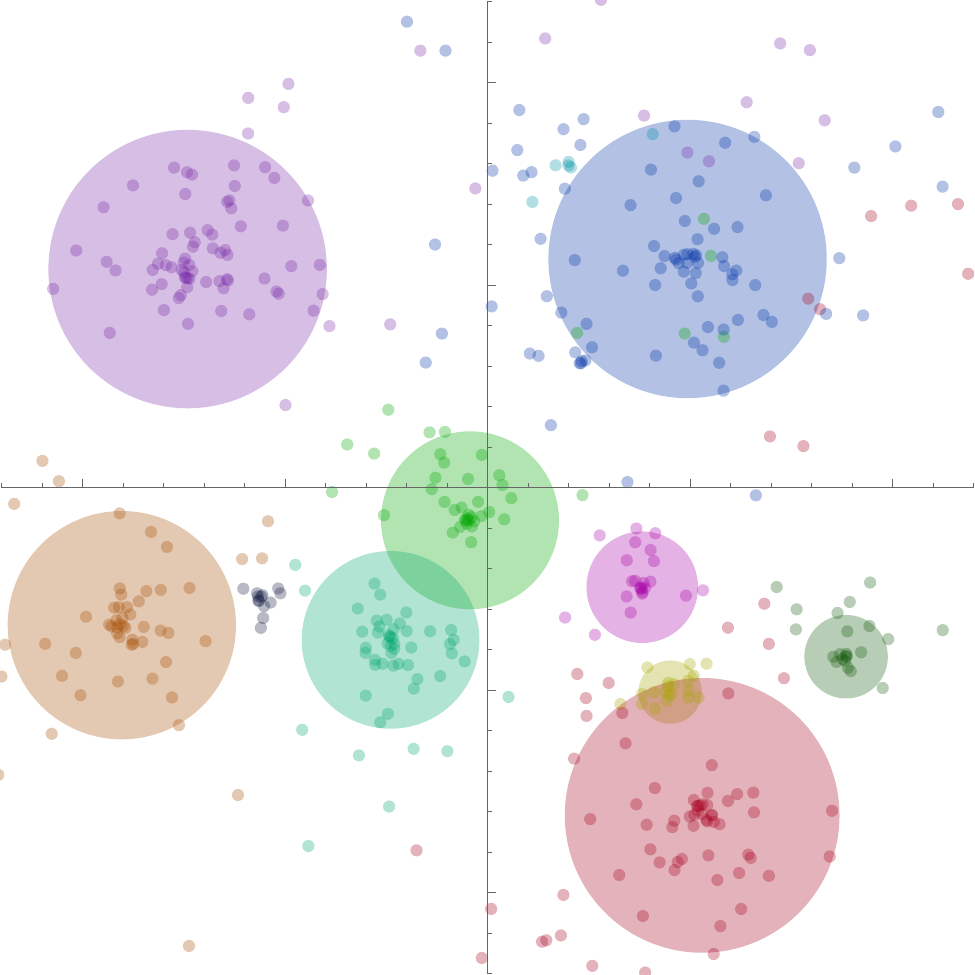}
\caption{Running example: second iteration.}
\label{fig:step2}
\end{figure}

With $\eta_j^k$ introduced, we define the affective shift vector:
\begin{equation}
    \psi_j^k := \begin{cases}
        \left(1 - \eta_j^k\right) \left(P_{V_j^k} - X_j^k\right) & \textnormal{if } A_j^k = 1 \textnormal{ or } V_j^k \neq W^k,\\
        (0,0) & \textnormal{if } A_j^k = 0 \textnormal{ and } V_j^k = W^k,
    \end{cases}
\end{equation}
reflecting the fact that for dissatisfied voters of the governing party, thermostatic effect overrides the affective shift.

Thermostatic effect is mechanically a mirror of the affective shift, with two exceptions: the shift is towards the first non-chosen party in the preference ordering, rather than the one previously chosen, and its magnitude is positively rather than negatively correlated with $\eta_j^k$ (as voters ideologically close to a party should be less likely to experience thermostatic shifts than those distant from it). Thus, the thermostatic shift equals
\begin{equation}
    \theta_j^k := \begin{cases}
        (0,0) & \textnormal{if } A_j = 1 \textnormal{ or } V_j^k \neq W^k,\\
        \frac{1}{2} \eta_j^k \left(P_{R_j^k} - X_j^k\right) & \textnormal{if } A_j = 0 \textnormal{ and } V_j^k = W^k,
    \end{cases}
\end{equation}
where
\begin{equation}
R_j^k = \operatorname*{arg\,max}_{i \in [n] \setminus \{W^k\}} U_{i,j}^k.
\end{equation}

Finally, we incorporate all adjustments into new voter positions: for every $k > 1$,
\begin{equation}
    X_j^k := X_j^{k-1} + \delta_j^{k-1} + \psi_j^{k-1} + \theta_j^{k-1}.
\end{equation}

States of an example realization of the model (voters and parties) after one, two, and ten iterations are plotted on, respectively, Figures \ref{fig:step1}, \ref{fig:step2}, and \ref{fig:step10}. For intervening steps of the same realization, see Appendix.

\subsection{Limitations}

Like many other computational methods, agent-based models involve a form of bias-variance tradeoff: as additional mechanisms are incorporated into the model, its complexity grows, leading to increases not only in technical features like Monte Carlo convergence times but also, more importantly, in fundamental characteristics such as the number of assumptions and free parameters or the risk of model overfitting. This tradeoff led us to forego a number of processes that, in our judgment, would not improve model accuracy sufficiently to justify that complexity cost. Most importantly:

\begin{itemize}
    \item We do not model any form of \emph{party system change}. In particular, our parties neither strategically change their ideal points to capture new voters nor shift in response to ideological, activist, and mobilization pressures; struggling parties neither merge nor fission; and new parties do not arise to capture unrepresented segments of the electorate. This is likely the most important limitation of our model. However, we believe that party system change does not pass the cost-benefit test under the bias-variance tradeoff. First, preliminary research suggests strong model sensitivity to the relative strength of the centripetal `strategic' push and the centrifugal `expressive' pull. Yet both are latent, and empirical calibration data are scarce. Second, evidence for ideological party polarization suggests that the `expressive' factor predominates. Thus, dropping party shifts altogether is the conservative choice: it likely understates polarization rather than risking overstatement.
    \item We do not model \emph{coalition formation} -- in our model, the largest party is always the governing party and bears all the costs of ruling. Again, there is insufficient evidence to choose between competing coalition formation models. Moreover, over three-fourths of coalitions are formed around the largest party \citep{Doring13}, and responsibility in coalition governments is usually attributed to the largest actor \citetext{\citealp{AngelovaEtAl16}; \citealp[but see][]{AllersEtAl22}}.
    \item Parties are perfectly homogeneous -- they always have a single ideal point -- while in reality they are collections of individuals with distinct ideal points whose cohesion sometimes breaks down. However, splitting parties into candidates and modeling cohesion would be computationally and conceptually costly, and cohesion is typically high enough that accuracy gains would be small.
    \item We do not model voter turnout, mobilization, and demobilization, assuming instead that all voters participate in all elections. This again is a conservative assumption, as it rules out ``base mobilization'' strategies, often cited as a driver of polarization.
    \item The set of voters does not change -- there is no influx of new voters, no disappearance of old voters. Given the largely stochastic nature of these processes, the accuracy cost of this simplification is likely minimal.
    % \item There is no long-term stable shift in social attitudes that would cause the center of gravity of the system to shift away from the original frame of reference.
\end{itemize}

\subsection{Modeling of electoral formulae}

For Jefferson--D'Hondt, we use the ``pot and ladle'' approximation formula \citep{FlisEtAl20,BoratynEtAl25b}. Mean district magnitude, $m$, is the parameter that determines the seat bias (which is proportional to $\hat{n} / (2m)$). The ``pot and ladle'' approximation is given by the following coordinatewise function:
\begin{equation}
    f_i((q_1, \dots, q_n)) = \begin{cases}
        \hat{q}_i \left(1 + \frac{\hat{n}}{2m}\right) - \frac{1}{2m} & \hat{q}_i \ge \frac{1}{2m + \hat{n}}, \\
        0 & \hat{q}_i < \frac{1}{2m + \hat{n}},
    \end{cases}
\end{equation}
where
\begin{equation}
    \hat{q}_i = \frac{q_i}{\sum_{l=1}^{\hat{n}} q_l^{\downarrow}},
\end{equation}
and
\begin{equation}
    \hat{n} = \max \left\{l = 1, \dots, n: \frac{q_l^{\downarrow}}{\sum_{i=1}^{l} q_i^{\downarrow}} \ge \frac{1}{2m + l}\right\}.
\end{equation}

For FPTP, we use the power law seats-votes formula \citep{Taagepera73,Tufte73,TaageperaShugart89}:
\begin{equation}
    f_i((q_1, \dots, q_n)) = \frac{q_i^\beta}{\sum_{l=1}^{n} q_l^\beta}.
\end{equation}
The power law exponent, $\beta$, is the parameter that controls how biased the electoral system is (the larger $\beta$, the more biased the system). In turn it corresponds to the concentration parameter of the spatial distribution of voters across districts.

\iffalse
\begin{figure}[htb]
\includegraphics[width=\columnwidth]{step6.pdf}
\caption{Running example: sixth iteration.}
\label{fig:step6}
\end{figure}
\fi

\begin{figure}[htb]
\includegraphics[width=\columnwidth]{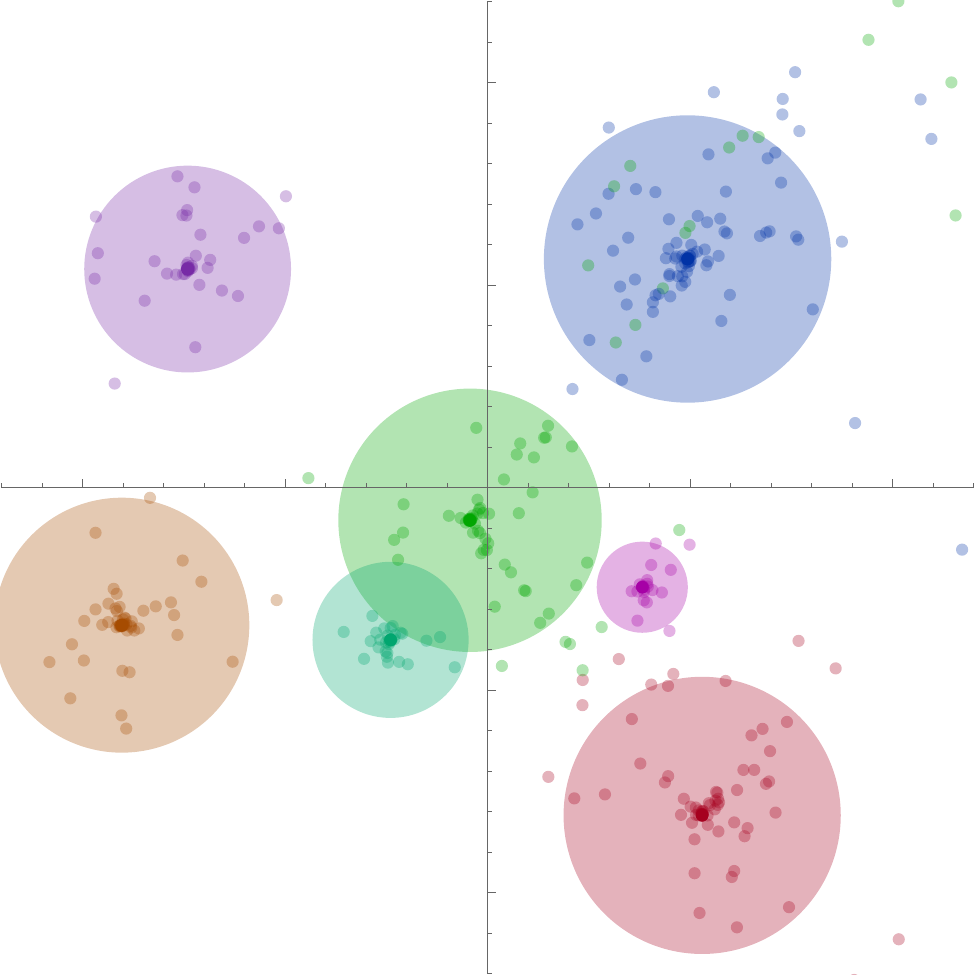}
\caption{Running example: last iteration.}
\label{fig:step10}
\end{figure}

\subsection{Measurement of polarization}

To measure voter polarization, we use the modified \citet{Mehlhaff24} index, defined as the proportion of inter-cluster variance of voter positions to their total variance. Because the Mehlhaff index is sensitive to the number of clusters (adding new clusters always increases inter-cluster variance, as it makes clusters smaller and therefore more cohesive), to make the result comparable we apply an ANOVA-style degrees-of-freedom correction, dividing the index by the number of degrees of freedom (number of clusters minus $1$). To identify the clusters involved, we use the \emph{mean shift clustering} algorithm with adaptive window size \citep{FukunagaHostetler75,ComaniciuMeer02,Cheng95}.

\section{Results}

For our primary simulation, we fixed the following hyperparameter values:
\begin{description}
    \item[$n = 12$] initial number of parties,
    \item[$r = 2$] radius of the disc containing party positions,
    \item[$\sigma = 1.5$] standard deviation of the distribution of initial voter positions,
    \item[$\rho = 2/3$] scaling factor for the distribution of `party base' voters,
    \item[$\mu = 2$] location parameter of the truncated normal distribution from which $\alpha_j$ is drawn,
    \item[$\tau = 0.25$] standard deviation parameter of the truncated normal distribution from which $\alpha_j$ is drawn,
    \item[$\pi = 0.125$] probability of a random perturbation of voter preferences,
    \item[$\varsigma = 0.25$] standard deviation of the distribution of random preference perturbations,
    \item[$\beta_1 = 15$] first parameter of the beta distribution from which the government job approval is drawn
    \item[$\beta_2 = 5$] second parameter of that distribution,
    \item[$\lambda = 5$] standard deviation parameter of the lognormal distribution from which voter shift parameter is drawn,
    \item[$v = 16384$] number of voters,
    \item[$k^{*} = 10$] number of iterations.
\end{description}

Some of those were chosen to match observable characteristics of election, such as the extent of strategic voting \citep{Spenkuch13} or the cost of ruling \citep{NannestadPaldam00}. Others were chosen by iterative experimentation to obtain realistic outcomes (as opposed to, say, a rapid collapse of the model towards a one-party rule).

\subsection{Model validity test}

\begin{figure}[htb]
\includegraphics[width=\columnwidth]{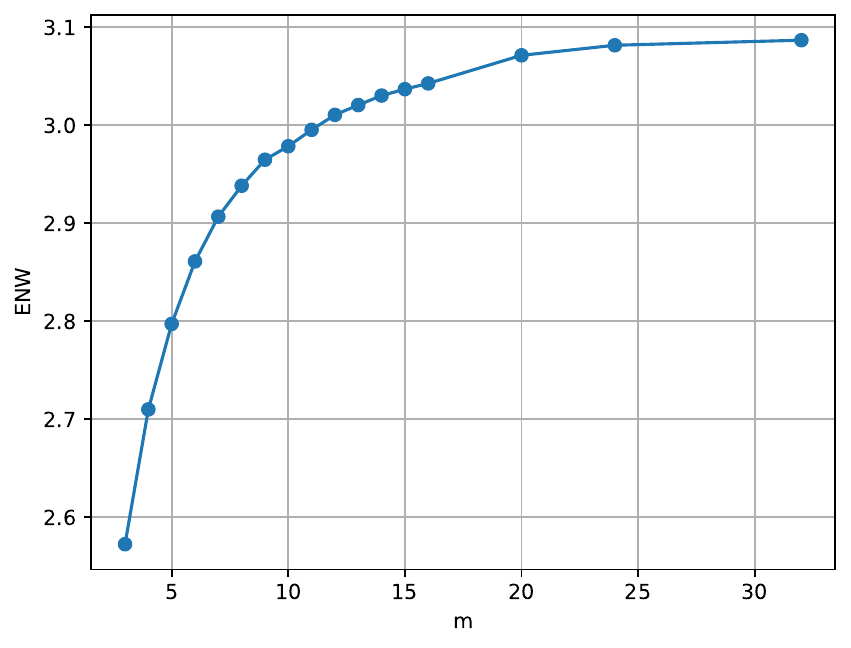}
\caption{Effective number of winners as a function of district magnitude ($m$). Note that smaller district magnitude implies greater seat bias in favor of larger parties.}
\label{fig:enw}
\end{figure}

\begin{figure}[tb]
\includegraphics[width=\columnwidth]{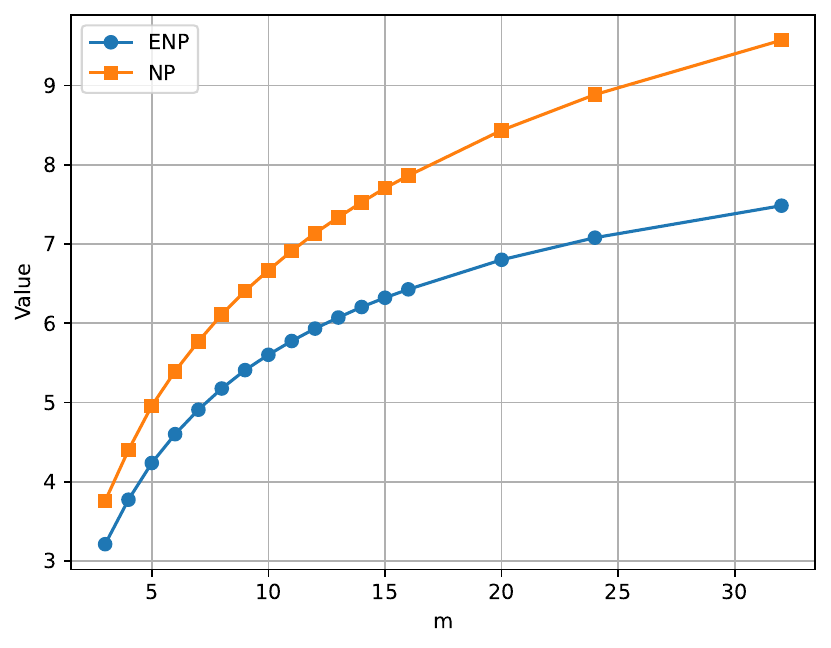}
\caption{Average number / effective number of parties at the end of the simulation as a function of district magnitude ($m$). Note that smaller district magnitude implies greater seat bias in favor of larger parties.}
\label{fig:enp}
\end{figure}

For each value of the district magnitude parameter $m = 3, \dots, 16, 20, 24, 32$, we have run 4096 simulations. Our first task is to assess whether the simulated party systems look substantively reasonable, before turning to the main question about polarization. To this end, we focus on three summary statistics computed at the end of each simulation run: the average number of parties that survive over all $10$ iterations, the average effective number of surviving parties, and the effective number of winners (to verify that the model generates meaningful alternation in office rather than trivial one–party dominance). The first two quantities are plotted in Figure~\ref{fig:enp}, while the effective number of winners is shown in Figure~\ref{fig:enw}.

Figures~\ref{fig:enp} and \ref{fig:enw} show that party-system size responds to district magnitude $m$ as theory predicts under Jefferson--D’Hondt \citep{FlisEtAl20}. At low $m$, strong large-party bias yields concentrated equilibria: few parties persist and the effective number of parties is low. As $m$ increases and seat bias weakens, both raw and effective party counts rise, indicating more fragmented multiparty competition; this is consistent with comparative evidence on magnitude and bias. Figure~\ref{fig:enw} provides a further plausibility check: the effective number of winners stays well above 1 (no permanent hegemony) yet does not grow without bound, implying alternation among a small set of governing parties rather than a fluid carousel. Most importantly, the outcomes appear broadly consistent with empirical patterns noted in countries employing the Jefferson--D'Hondt system, such as Poland, Czech Republic, Finland, Portugal, Croatia, or Luxembourg (Table \ref{tbl:countries}).

\begin{table}[htb]
\begin{tabular}{lrrrrr}
\toprule
\textbf{country} & \textbf{mean $m$} & \textbf{mean $n$} & \textbf{RP} & \textbf{ENP} & \textbf{ENW} \\
\midrule
Poland           & 11.22      & 10.23      & 5.88         & 4.95         & 3.5 \\
Czech Republic   & 14.29      & 10.56      & 6.33         & 5.85         & 3.57 \\
Finland          & 13.79      & 9.23       & 6.77         & 5.37         & 2.78 \\
Portugal         & 10.74      & 7.53       & 4.65         & 3.47         & 2.63 \\
Croatia          & 13.09      & 13.57      & 7.86         & 4.38         & 1.47 \\
Spain            & 6.73       & 13.47      & 7.13         & 4.13         & 1.92 \\
Luxembourg       & 14.43      & 6.88       & 4.21         & 3.89         & 1.47 \\
Netherlands      & 145.45     & 11.55      & 7.90         & 5.60         & 3.57 \\ \bottomrule \smallskip
\end{tabular}
\caption{EU countries employing Jefferson--D'Hondt: average district magnitude ($m$), average number of national parties ($m$), average number of represented parties (RP), average effective number of parties (ENP), and effective number of winners (ENW).}
\label{tbl:countries}
\end{table}

\subsection{Hyperparameter sensitivity}

The next preliminary question concerns model sensitivity to specific parameter values. For every major hyperparameter, we have varied its value (usually from 50\% to 200\% of the initial value) and examined how those changes affect the number of surviving parties, the effective number of parties, and the effective number of winners.

\begin{figure}[htb]
\includegraphics[width=\columnwidth]{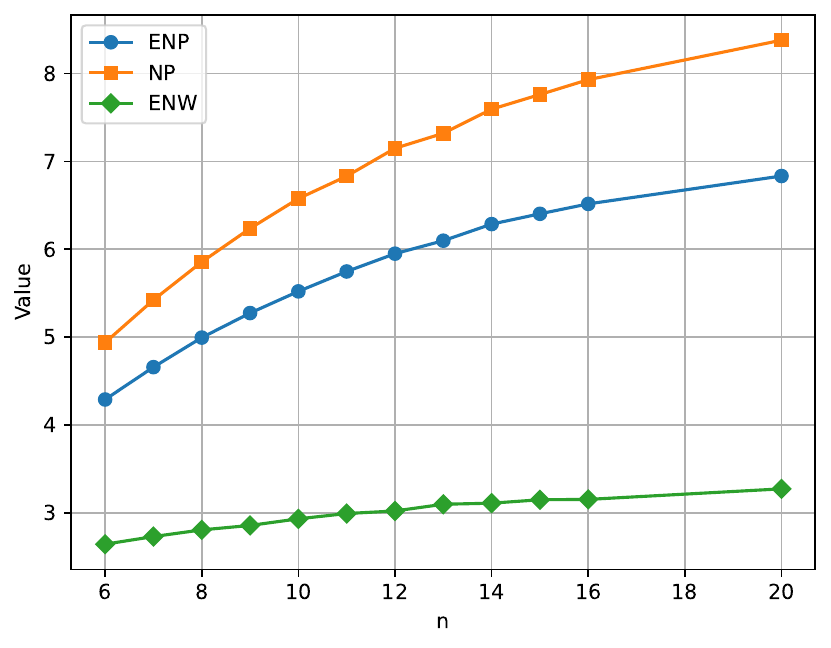}
\caption{Sensitivity of model outputs to variation in $n$.}
\label{fig:sensN}
\end{figure}

\begin{figure}[htb]
\includegraphics[width=\columnwidth]{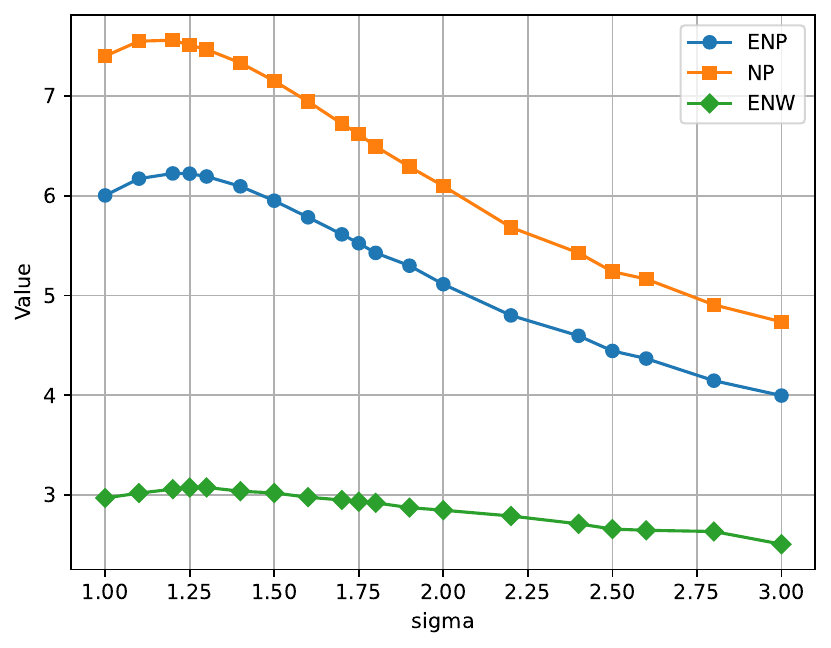}
\caption{Sensitivity of model outputs to variation in $\sigma$ (initial voter dispersion).}
\label{fig:sensSigma}
\end{figure}

\begin{figure}[htb]
\includegraphics[width=\columnwidth]{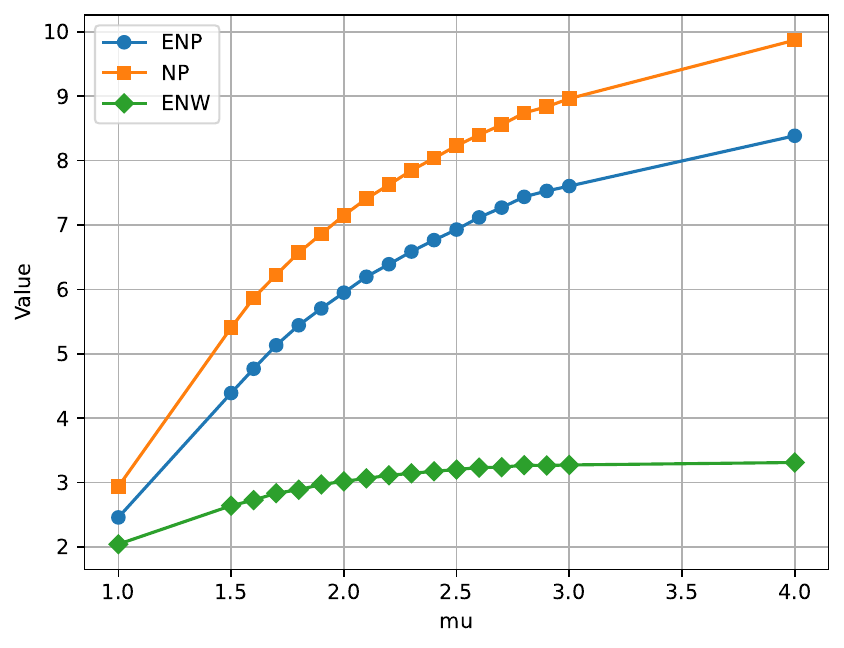}
\caption{Sensitivity of model outputs to variation in $\mu$ (strategy-proximity tradeoff expectation).}
\label{fig:sensMu}
\end{figure}

\begin{figure}[htb]
\includegraphics[width=\columnwidth]{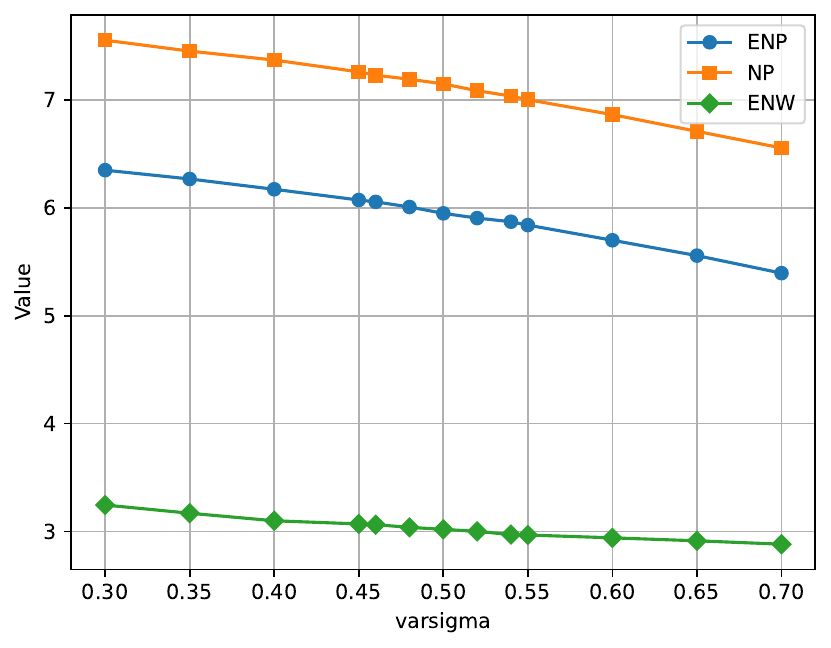}
\caption{Sensitivity of model outputs to variation in $\varsigma$ (affective shift variance).}
\label{fig:sensVarSigma}
\end{figure}

\begin{figure}[htb]
\includegraphics[width=\columnwidth]{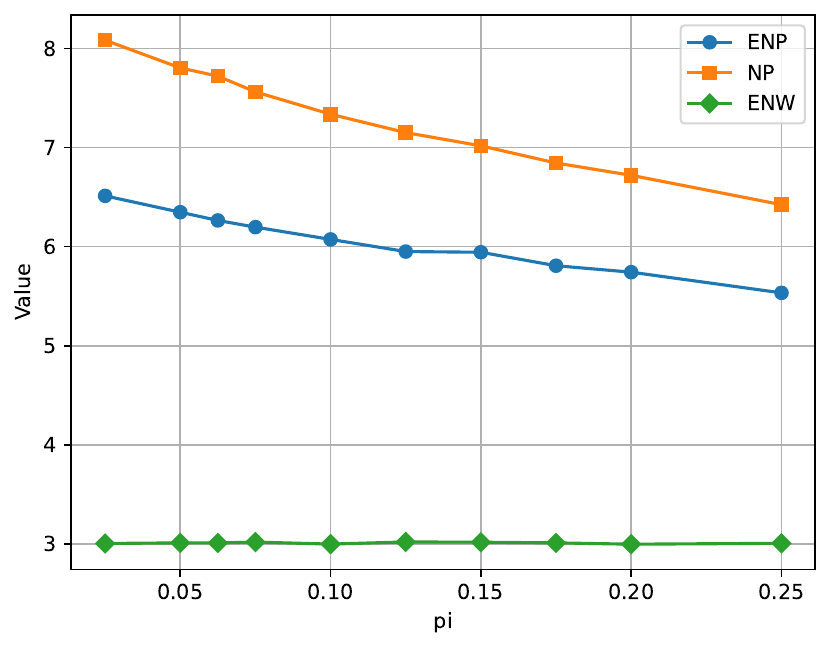}
\caption{Sensitivity of model outputs to variation in $\pi$ (random shift probability).}
\label{fig:sensPi}
\end{figure}

Model outputs are most sensitive to changes in the initial number of parties ($n$, see Figure \ref{fig:sensN}), $\sigma$ (Figure \ref{fig:sensSigma}), and $\mu$ (Figure \ref{fig:sensMu}). In the case of the number of parties, the effect is quite expected: starting with fewer parties should naturally lead to fewer surviving parties and less power alternation. Similarly, neither positive correlation between party system fragmentation and voter emphasis on policy proximity, which translating into weaker `gravitational shift' of large parties (Figure \ref{fig:sensMu}), nor negative correlation between fragmentation and affective shift variance (Figure \ref{fig:sensVarSigma}) is particularly surprising. On the other hand, the effect for $\sigma$ is quite surprising: apparently, the more dispersed the electorate at the outset, the less fragmented the party system becomes $\sigma$ (Figure \ref{fig:sensSigma}). This is likely due to voter dispersion weakening centrist parties and accelerating their disappearance.

\subsection{Seat bias and polarization -- Jefferson--D'Hondt}

\begin{figure}[htb]
\includegraphics[width=\columnwidth]{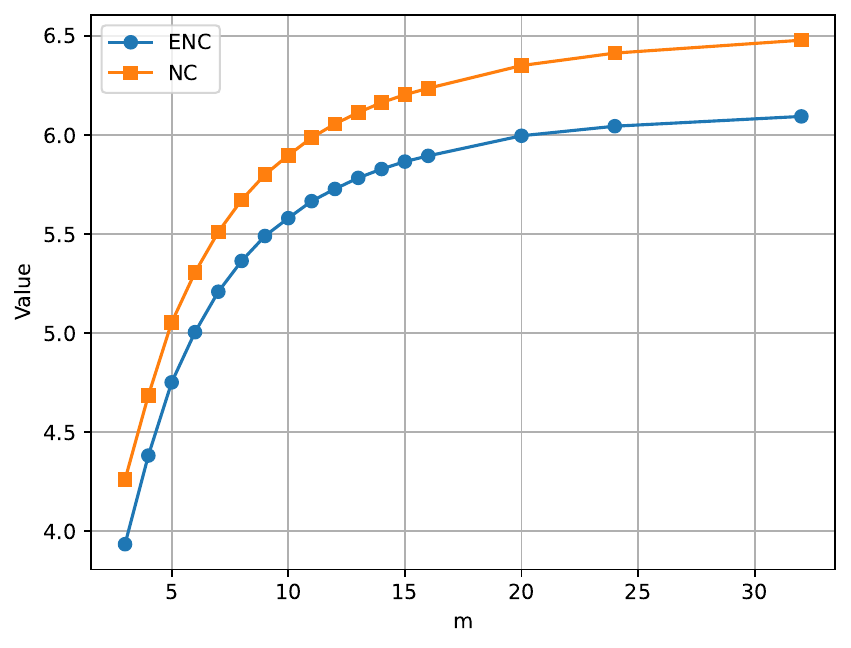}
\caption{Average number / effective number of voter clusters at the end of the simulation as a function of district magnitude ($m$). Note that smaller district magnitude implies greater seat bias in favor of larger parties.}
\label{fig:enc}
\end{figure}

We now turn to our main outcome of interest: voter polarization. Figure~\ref{fig:enc} plots the average number and effective number of voter clusters at the end of the simulation as a function of district magnitude. The figure shows that under strongly seat-biased conditions (low $m$), voters tend to sort into a relatively small number of cohesive clusters. As district magnitude increases and the electoral system becomes more permissive toward smaller parties, additional clusters emerge and existing ones become somewhat less sharply delineated. This pattern is consistent with the intuition that highly biased systems encourage voters to coordinate on a few large poles of competition, while more proportional systems sustain a more differentiated, but also less sharply separated, set of political groupings.

\begin{figure}[htb]
\includegraphics[width=\columnwidth]{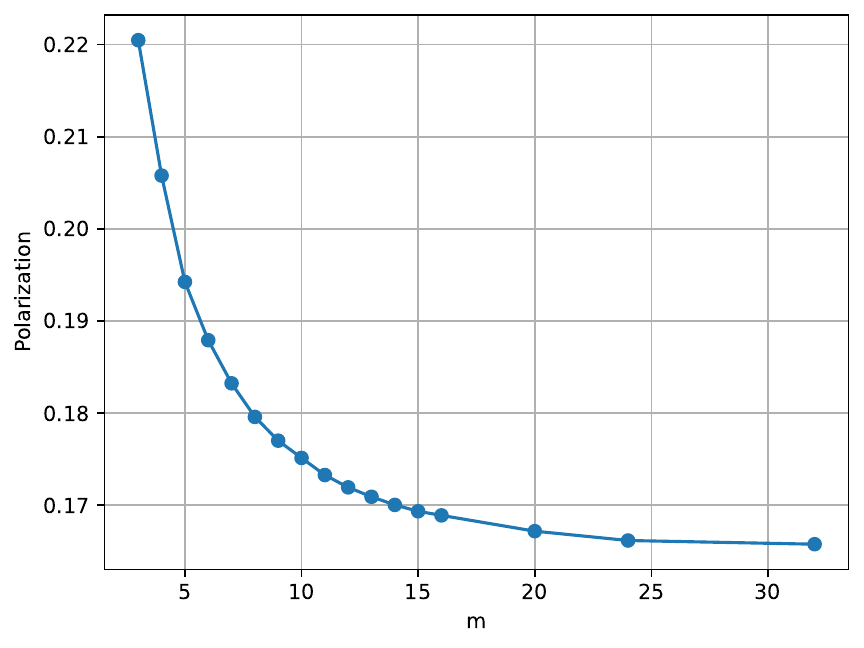}
\caption{Polarization as a function of district magnitude ($m$). Note that smaller district magnitude implies greater seat bias in favor of larger parties.}
\label{fig:polar}
\end{figure}

Figure~\ref{fig:polar} summarizes these changes using our polarization index, which captures the proportion of total variance in voter positions that is attributable to between-cluster differences (with a degrees-of-freedom correction). Visually, the relationship between district magnitude and polarization is clearly negative: simulations run under low-$m$ configurations exhibit substantially higher values of the polarization index than those run under high-$m$ configurations. In other words, when the electoral system systematically favors larger parties, the voter distribution in policy space evolves toward more clearly separated clusters than under less biased systems.

% \subsection{Hypothesis testing}

\begin{table}[htbp]
\centering
\begin{tabular}{lrrrr}
\hline
 & Estimate & Std. Error & t value & Pr(>$|t|$) \\
\hline
(Intercept) & 0.2030 & 0.0052 & 39.03 & 1.09 E-15 \\
$m$         & -0.0021 & 0.00042 & -5.02 & 1.87 E-04 \\
\hline
\multicolumn{5}{l}{Residual standard error: 0.00945 on 14 df} \\
\multicolumn{5}{l}{Multiple $R^2$: 0.6431,\quad Adjusted $R^2$: 0.6176} \\
\multicolumn{5}{l}{F-statistic: 25.23 on 1 and 14 df,\quad $p$-value: $1.87\times 10^{-4}$} \\
\hline
\end{tabular}
\caption{Linear regression of polarization on district magnitude ($m$).}
\label{tbl:reg}
\end{table}

Table~\ref{tbl:reg} reports a simple linear regression of average polarization on district magnitude. The estimated intercept is $0.2030$ and the slope is $-0.00210$, with a $p$-value of $1.87 \times 10^{-4}$. Substantively, this implies that increasing $m$ by one unit is associated, on average, with a decrease in the polarization index of about $0.0021$ points. Over the range of $m$ values that we consider (from $3$ to $32$), this corresponds to a reduction in predicted polarization of roughly $0.06$ points, i.e., over one-fourth of the observed maximum. The model explains approximately $64\%$ of the variance in average polarization across the different values of $m$ ($R^2 = 0.64$) despite relatively poor fit (the curve on Figure~\ref{fig:polar} is decidedly not affine). While the functional form is deliberately kept simple, these results confirm what inspection of Figure~\ref{fig:polar} already suggests: in our simulations, stronger large-party seat bias (lower $m$) is systematically associated with higher long-run polarization among voters.

\subsection{Seat bias and polarization -- FPTP (plurality)}

\begin{figure}[htb]
\includegraphics[width=\columnwidth]{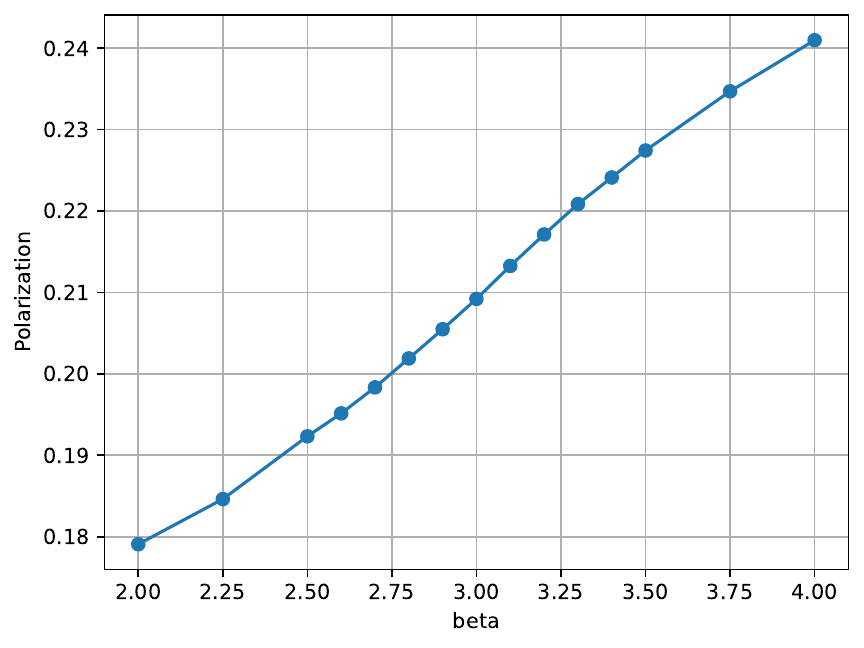}
\caption{Polarization as a function of the power law exponent ($\beta$). Note that larger exponent implies more concentrated geographic vote distribution, and thus higher bias.}
\label{fig:fptpPolar}
\end{figure}

The hypothesized relation between bias and polarization is also confirmed for the first-past-the-post (plurality) rule. Figure \ref{fig:fptpPolar} plots how the polarization index changes with the power law exponent (for most real-life FPTP systems, $\beta \approx 3$). Notably, low-bias plurality systems can actually be less polarizing than high-bias Jefferson--D'Hondt systems, though this may be an artifact of the power law approximation lacking a natural threshold while the ``pot-and-ladle'' approximation has one. Visual examination strongly suggest that the positive association between bias and polarization persists for plurality systems, and this conclusion is confirmed by another linear regression model (Table~\ref{tbl:reg2}). The slope of polarization on $\beta$ is $0.0332$, with a $p$-value of $1.46 \times 10^{-15}$. The model explains nearly all of the variance in average polarization across the different values of $m$ ($R^2 > 0.99$).

\begin{table}[htbp]
\centering
\begin{tabular}{lrrrr}
\hline
 & Estimate & Std. Error & t value & Pr(>$|t|$) \\
\hline
(Intercept) & 0.1099 & 0.0023 & 48.04 & 5.03 E-16 \\
$\beta$     & 0.0332 & 0.00075 & 44.24 & 1.46 E-15 \\
\hline
\multicolumn{5}{l}{Residual standard error: 0.001544 on 13 df} \\
\multicolumn{5}{l}{Multiple $R^2$: 0.9934,\quad Adjusted $R^2$: 0.9929} \\
\multicolumn{5}{l}{F-statistic: 1957 on 1 and 13 df,\quad $p$-value: $1.46\times 10^{-15}$} \\
\hline
\end{tabular}
\caption{Linear regression of polarization on power law coefficient ($\beta$).}
\label{tbl:reg2}
\end{table}

\subsection{Robustness}

We have also investigated the robustness of these findings with respect to changes in the hyperparameters of the model. For each parameter in turn, we varied its value within a $\pm 50\%$ band around the baseline specification and re-estimated the relationship between district magnitude (or power-law coefficient, as the case may be) and polarization. Although these perturbations do affect the overall level of polarization (for example, stronger affective shifts or weaker random perturbations tend to raise the index across all institutional settings), they do not alter the sign of the association between $m$ (or $beta$) and polarization. In all cases examined, configurations with lower district magnitude (and hence greater seat bias) yield higher average polarization than otherwise identical configurations with higher $m$ or $\beta$.

\subsection{Conclusions}

Within the confines of our modeling assumptions, these results support a clear conclusion: electoral systems that confer larger seat bonuses on large parties tend to generate more polarized voter distributions over repeated electoral cycles. Seat bias is therefore not merely a question of mechanical disproportionality; in our simulations, it acts as a substantive driver of political polarization by encouraging voters to cluster around a small number of distinct poles in policy space. While our model is deliberately stylized and does not claim to capture every feature of real-world electorates, it provides strong evidence that large-party seat bias is a significant contributing factor to the increase of political polarization.

\section{Acknowledgments}

This work has been supported by the Polish National Center for Science (NCN) under grants no. 2019/35/B/HS5/03949 (\textit{Political Effects of Select Electoral Systems from the Quantitative Perspective}) and 2023/49/B/HS5/03893 (\textit{Voter Shifts and Spatial Models of Party Competition}). Computational resources were provided by the Jagiellonian University Excellence Initative (QuantPol project).

\FloatBarrier

% References
\bibliography{polar}

%%%%%%%%%%%%%%%%%%%%%%%%%%%%%%%%%%%%%%%%%%%%%%%%%%%%%%%%%%%%%%%%%%%%%%%%%

\clearpage
\section{Appendix}

\begin{figure}[H]
\includegraphics[width=0.85\columnwidth]{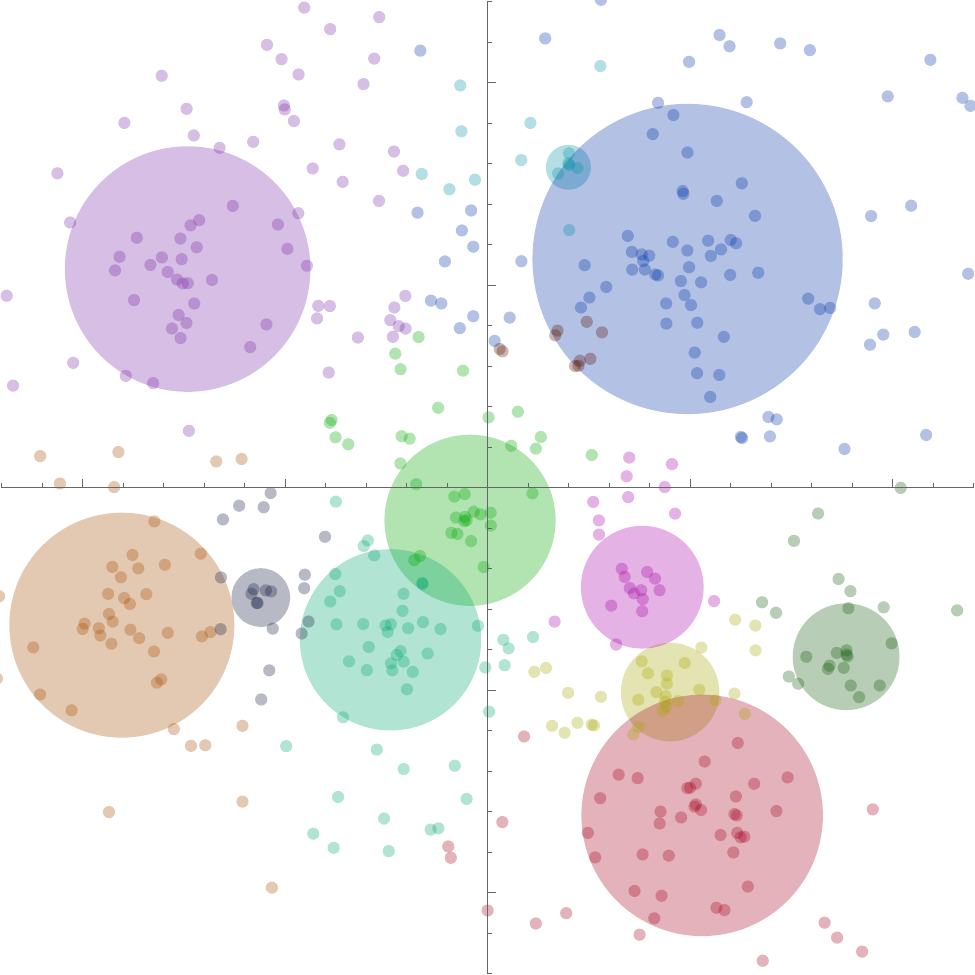}
\caption{Running example: first iteration.}
\label{fig:step1a}
\end{figure}

\begin{figure}[H]
\includegraphics[width=0.85\columnwidth]{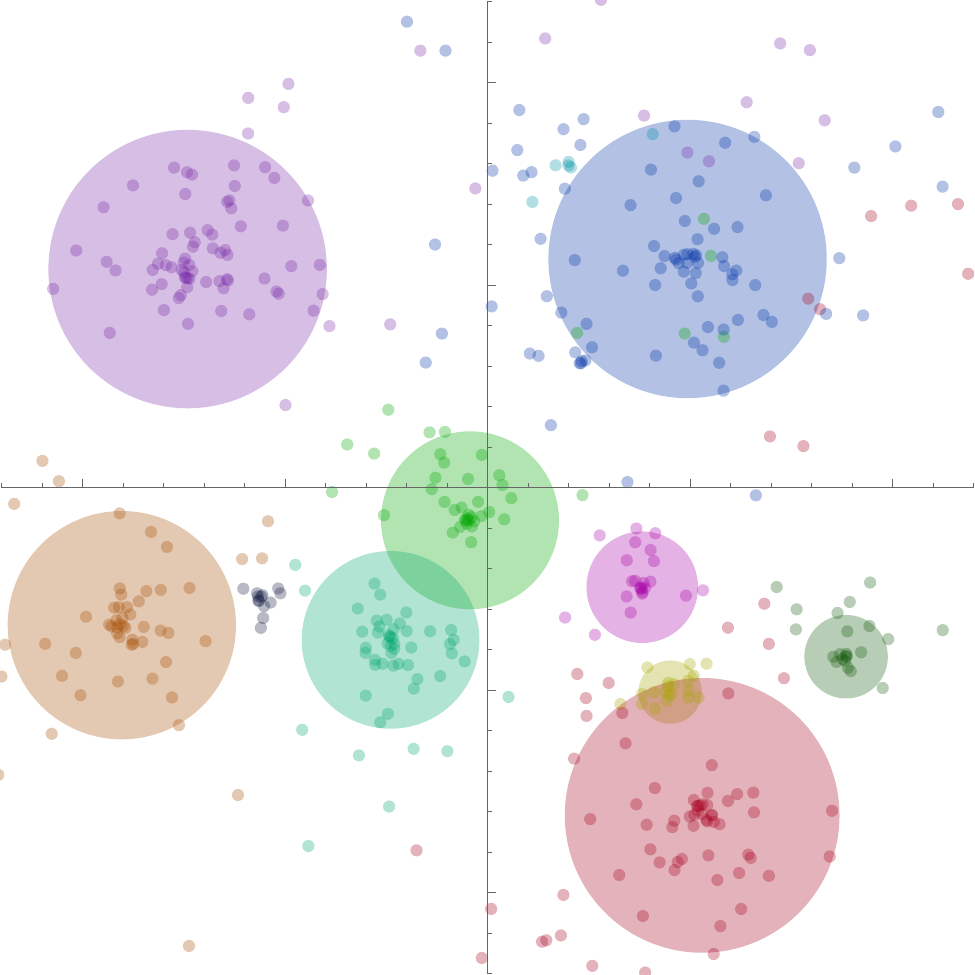}
\caption{Running example: second iteration.}
\label{fig:step2a}
\end{figure}

\begin{figure}[H]
\includegraphics[width=0.85\columnwidth]{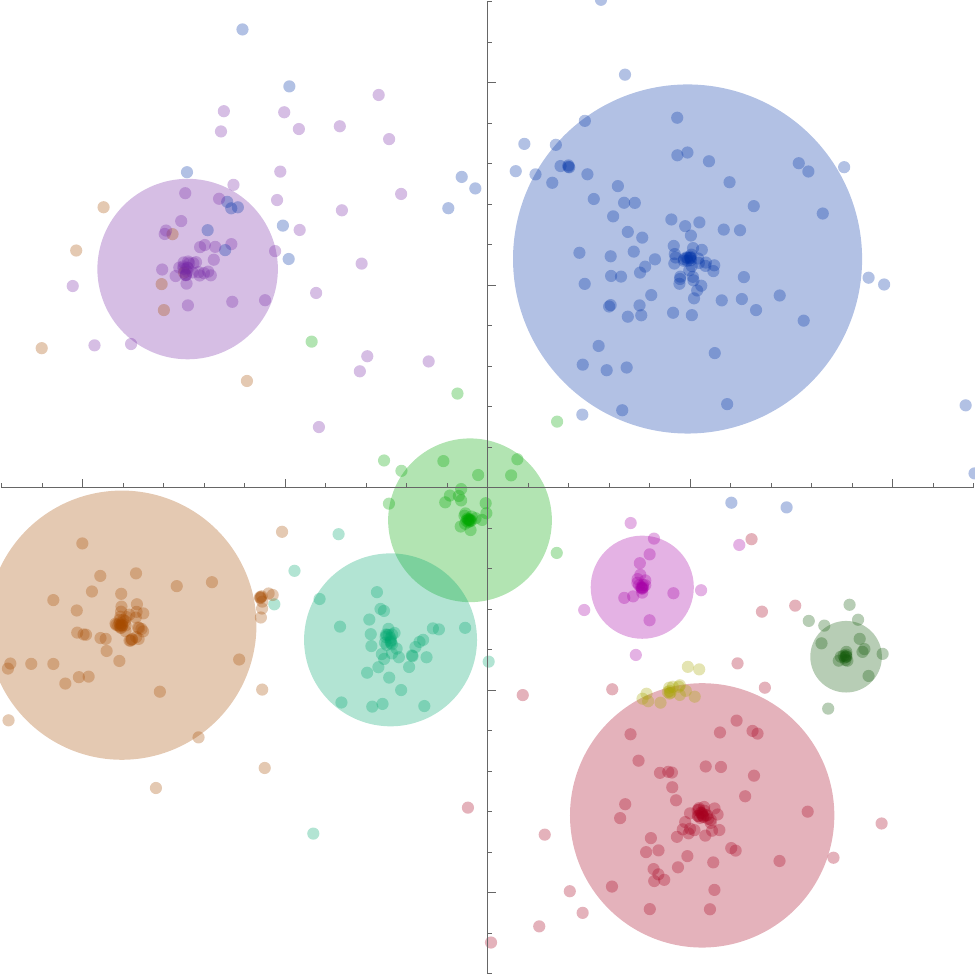}
\caption{Running example: third iteration.}
\label{fig:step3}
\end{figure}

\vspace{-1.5cm}
\begin{figure}[H]
\includegraphics[width=0.85\columnwidth]{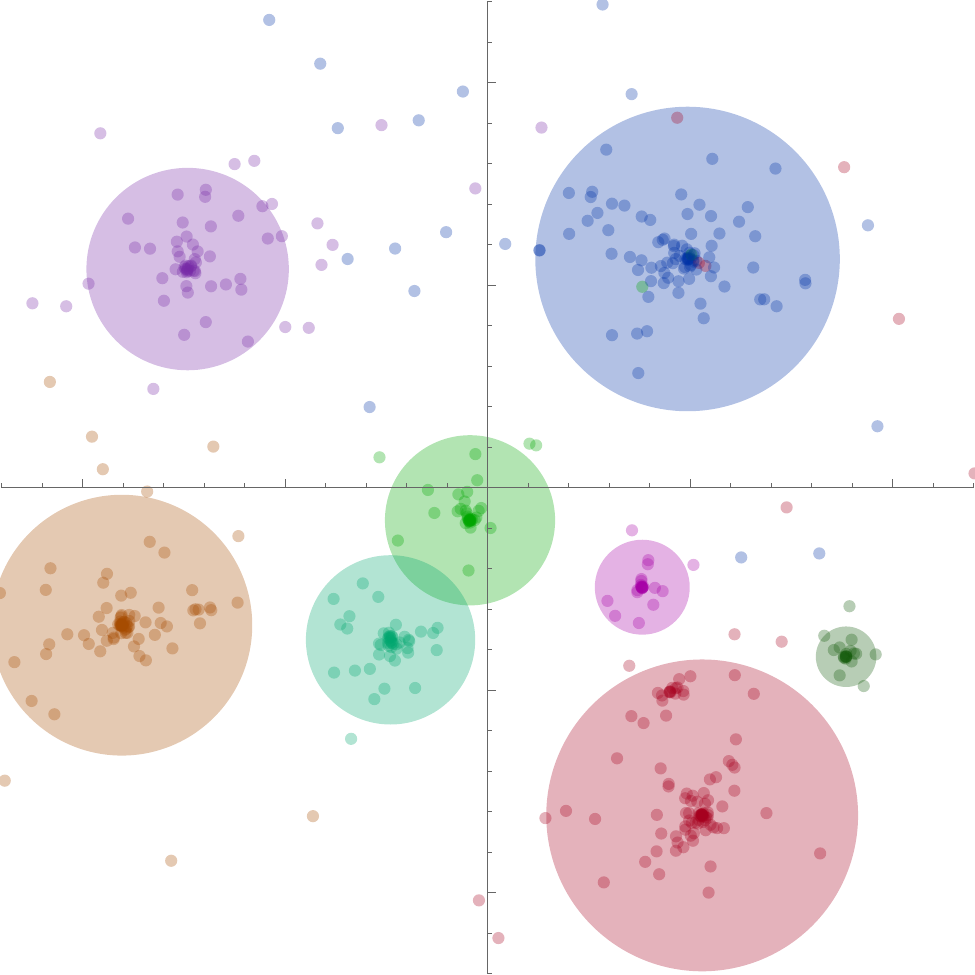}
\caption{Running example: fourth iteration.}
\label{fig:step4}
\end{figure}

\vspace{-1.5cm}
\begin{figure}[H]
\includegraphics[width=0.85\columnwidth]{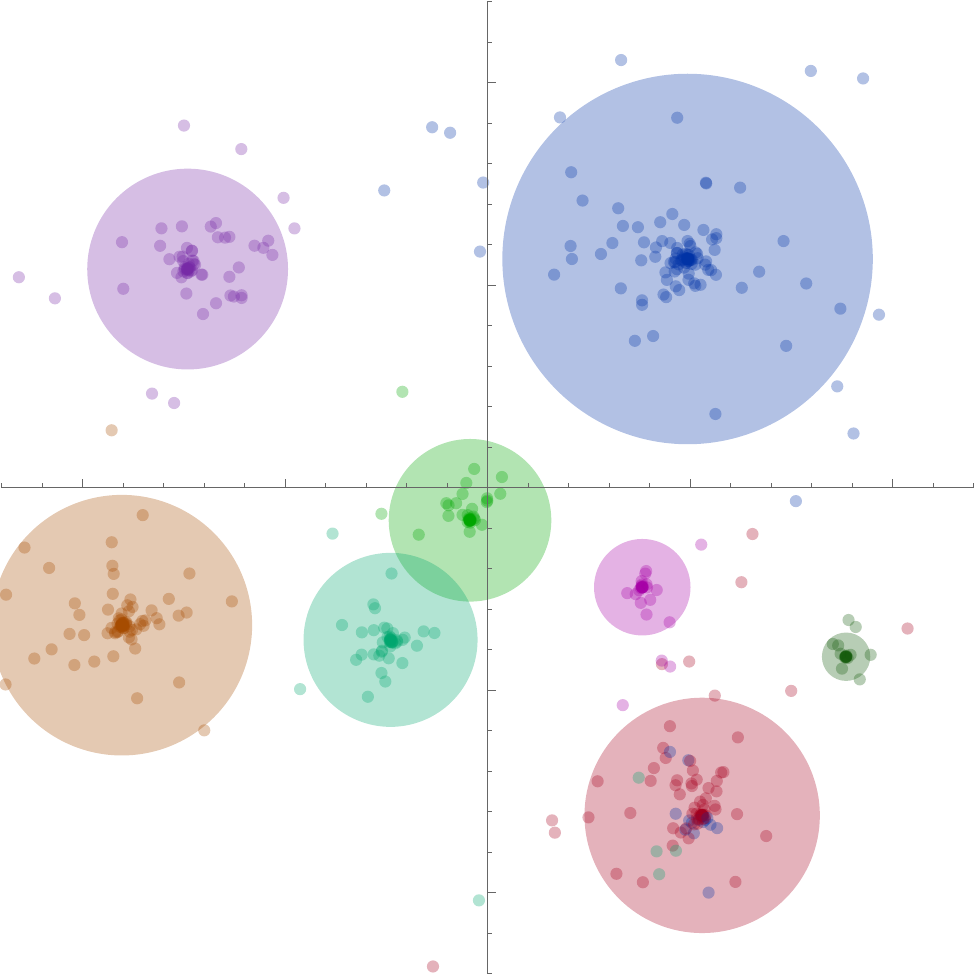}
\caption{Running example: fifth iteration.}
\label{fig:step5}
\end{figure}

\begin{figure}[H]
\includegraphics[width=0.85\columnwidth]{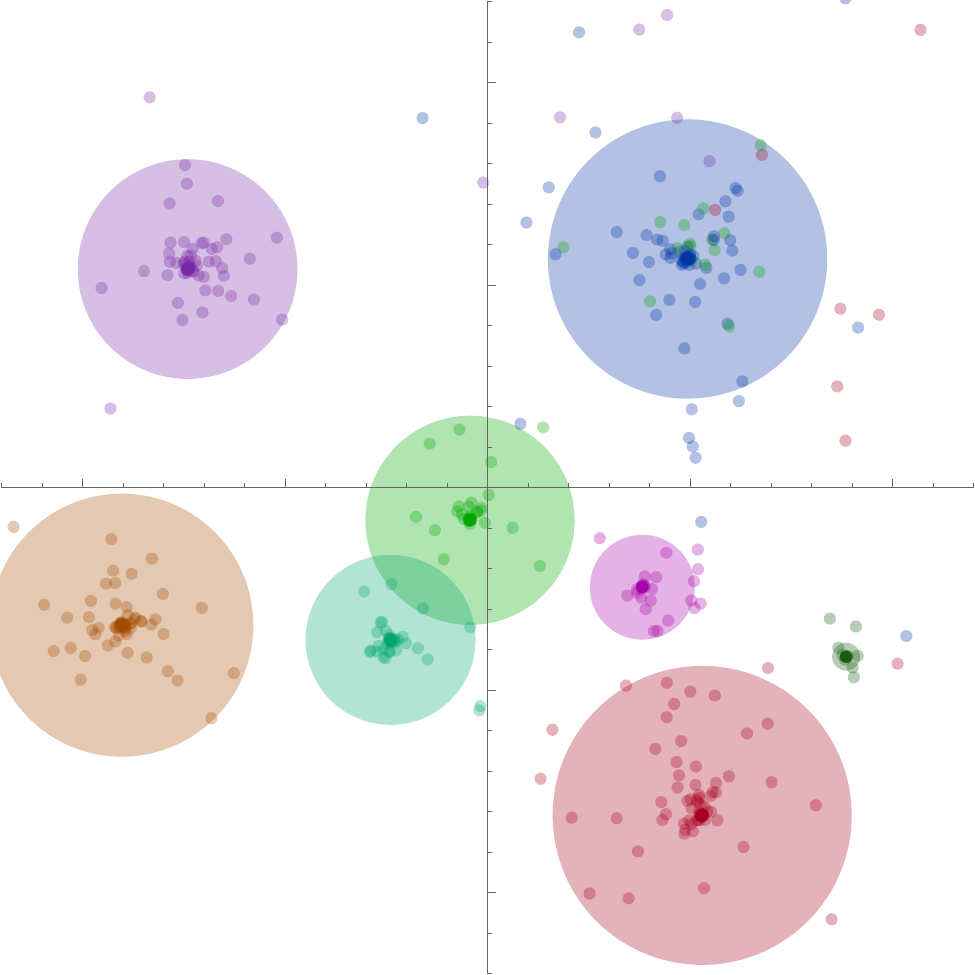}
\caption{Running example: sixth iteration.}
\label{fig:step6}
\end{figure}

\vspace{-1.5cm}
\begin{figure}[H]
\includegraphics[width=0.85\columnwidth]{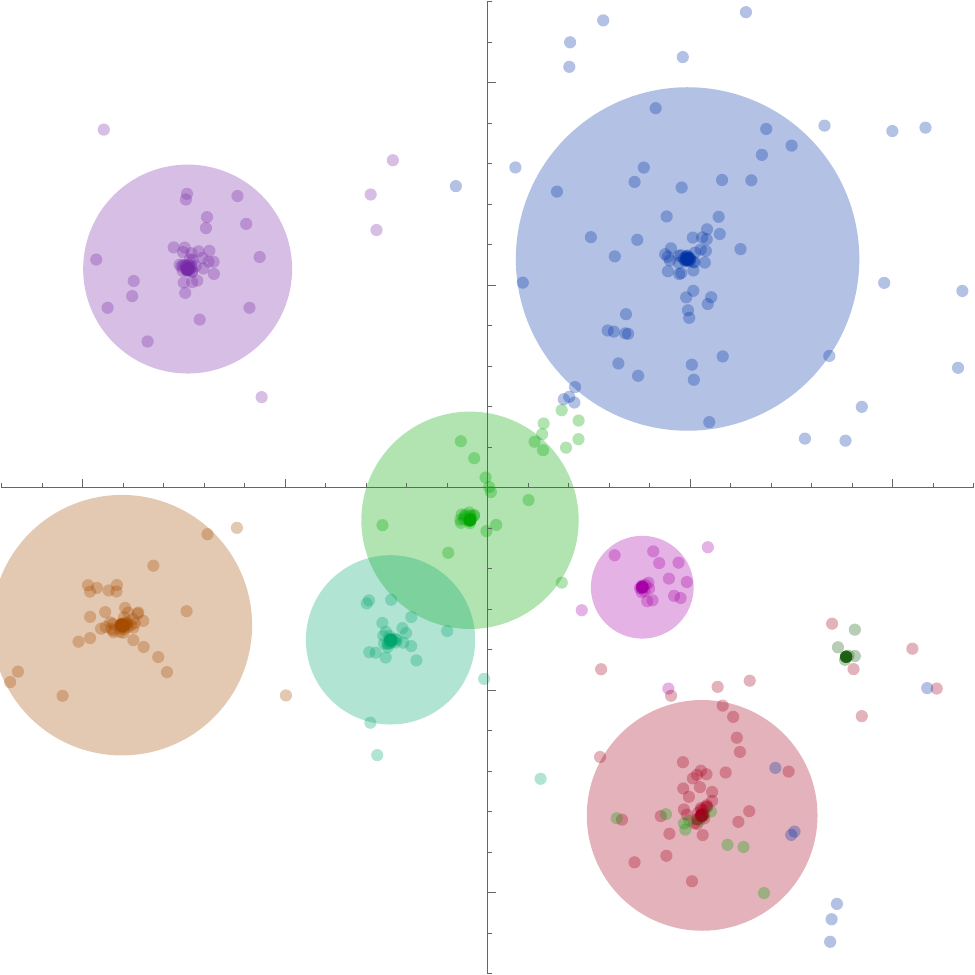}
\caption{Running example: seventh iteration.}
\label{fig:step7}
\end{figure}

\vspace{-1.5cm}
\begin{figure}[H]
\includegraphics[width=0.85\columnwidth]{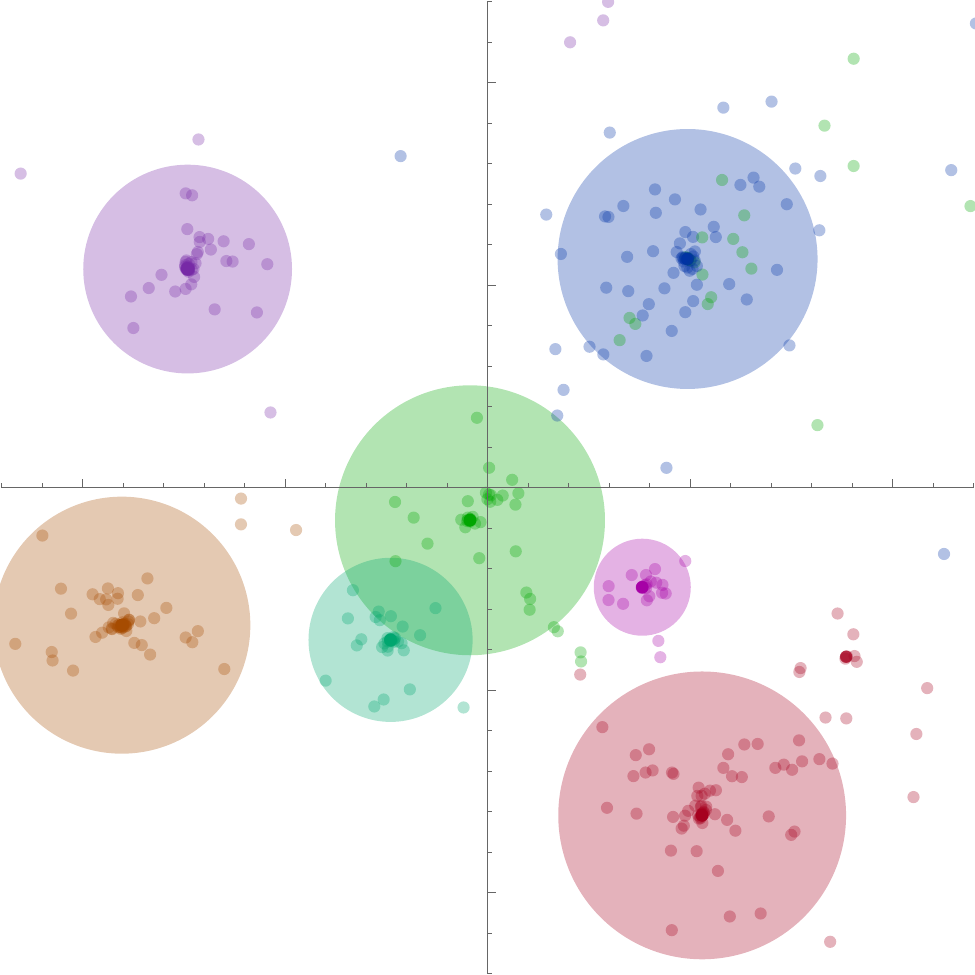}
\caption{Running example: eighth iteration.}
\label{fig:step8}
\end{figure}

\begin{figure}[H]
\includegraphics[width=0.85\columnwidth]{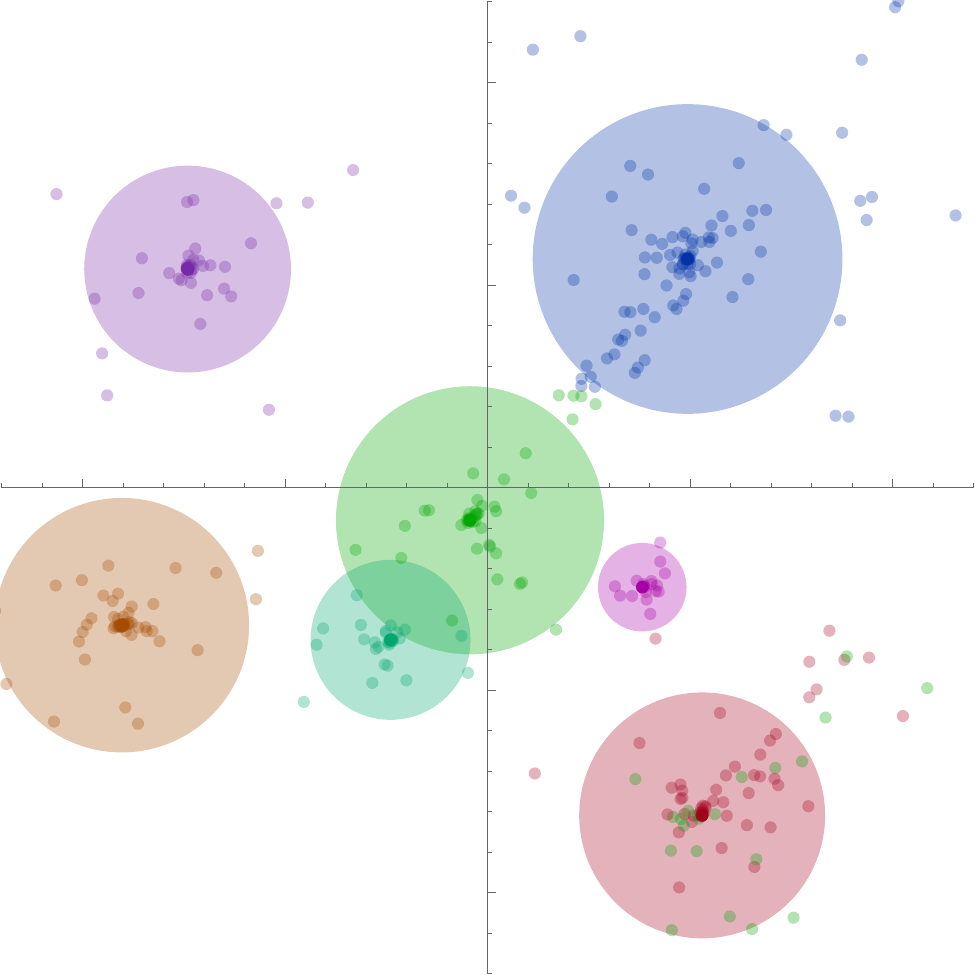}
\caption{Running example: ninth iteration.}
\label{fig:step9}
\end{figure}

\begin{figure}[H]
\includegraphics[width=0.85\columnwidth]{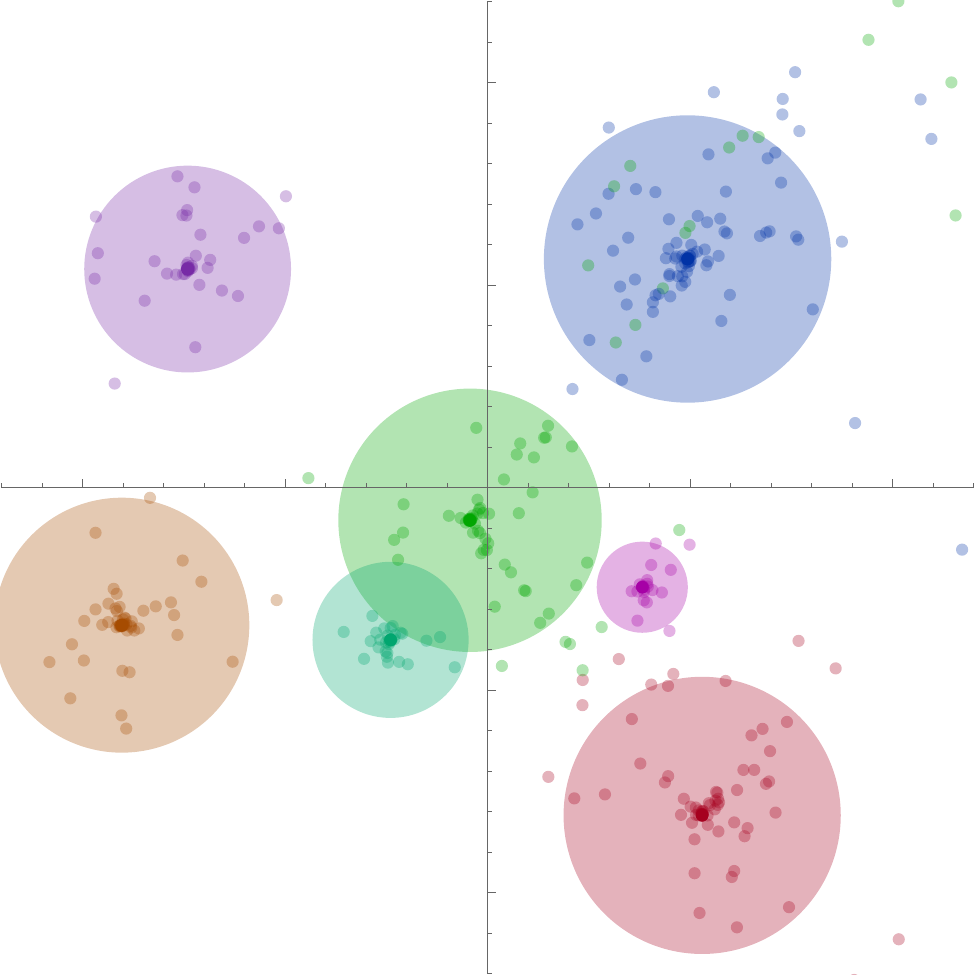}
\caption{Running example: tenth iteration.}
\label{fig:step10a}
\end{figure}

\end{document}